\documentclass[aps,reprint,superscriptaddress,longbibliography]{revtex4-2}
\usepackage[T1]{fontenc}
\usepackage{physics}
\usepackage{mathtools}
\usepackage{amsmath}
\usepackage{amssymb}
\usepackage{amsthm}
\usepackage{algorithm}
\usepackage{algpseudocode}
\usepackage[usenames,dvipsnames]{xcolor} 
\usepackage[colorlinks=true,citecolor=Purple,linkcolor=Purple,urlcolor=Purple
]{hyperref}
\usepackage{dsfont}
\usepackage{enumitem}
\usepackage{cases}
\usepackage{float}
\usepackage{subfigure}
\usepackage{dcolumn}
\usepackage{bm}
\usepackage{svg} 
\usepackage{xpatch}
\usepackage{multirow}
\usepackage{array, makecell}
\makeatletter
\xpatchcmd{\@ssect@ltx}{\@xsect}{\protected@edef\@currentlabelname{#8}\@xsect}{}{}
\xpatchcmd{\@sect@ltx}{\@xsect}{\protected@edef\@currentlabelname{#8}\@xsect}{}{}
\makeatother

\newcommand{\eq}[1]{(\ref{eq:#1})}

\newcommand{\calL}{\mathcal{L}} 
 
\newcommand{\x}{\boldsymbol{x}} 
\newcommand{\bt}{\boldsymbol{\theta}}

\newcommand{\thv}{{\boldsymbol \theta}}

\newcommand{\HC}{\mathcal{H}}
\newcommand{\LC}{\mathcal{L}}
\newcommand{\OC}{\mathcal{O}}
\newcommand{\TC}{\mathcal{T}}

\newcommand{\la}[1]{{\color{red}#1}}
\newcommand{\LA}[1]{{\color{red}$\big[\![$~\raisebox{.7pt}{L}\!\!\:\raisebox{-.7pt}{A}: \textit{#1}~$]\!\big]$}}

\begin{document}

\preprint{APS/123-QED}

\title{Towards large-scale quantum optimization solvers with few qubits}

\author{Marco Sciorilli$^{\dag}$}
\thanks{Both authors contributed equally to this work}
\affiliation{Quantum Research Center, Technology Innovation Institute, Abu Dhabi, UAE}%
\email{Marco.Sciorilli@tii.ae}

\author{Lucas Borges}
\thanks{Both authors contributed equally to this work}
\affiliation{Quantum Research Center, Technology Innovation Institute, Abu Dhabi, UAE}
\affiliation{Federal University of Rio de Janeiro, Caixa Postal 652, Rio de Janeiro, RJ 21941-972, Brazil}%

\author{Taylor L. Patti}
\affiliation{NVIDIA, Santa Clara, California 95051, USA}
\author{Diego Garc\'ia-Mart\'in}
\affiliation{%
Information Sciences, Los Alamos National Laboratory, Los Alamos, NM 87545, USA
}%
\affiliation{Quantum Research Center, Technology Innovation Institute, Abu Dhabi, UAE}
\author{Giancarlo \surname{Camilo}}
\affiliation{Quantum Research Center, Technology Innovation Institute, Abu Dhabi, UAE}
\author{Anima \surname{Anandkumar}}
\affiliation{Department of Computing + Mathematical Sciences (CMS), California Institute of Technology (Caltech), Pasadena, CA, 91125 USA}
\author{Leandro \surname{Aolita}}
\affiliation{Quantum Research Center, Technology Innovation Institute, Abu Dhabi, UAE}

\date{\today}

\begin{abstract}
We introduce a variational solver for combinatorial optimizations over $m=\order{n^k}$ binary variables using only $n$ qubits, with tunable $k>1$. 
The number of parameters and circuit depth display mild linear and sublinear scalings in $m$, respectively.
Moreover, we analytically prove that the specific qubit-efficient encoding brings in a super-polynomial mitigation of barren plateaus as a built-in feature. 
This leads to unprecedented quantum-solver performances. 
For $m=7000$, numerical simulations produce solutions competitive in quality with state-of-the-art classical solvers. 
In turn, for $m=2000$, experiments with $n=17$ trapped-ion qubits feature MaxCut approximation ratios estimated to be beyond the hardness threshold $0.941$. 
To our knowledge, this is the highest quality attained experimentally on such sizes. 
Our findings offer a novel heuristics for quantum-inspired solvers as well as a promising route towards solving commercially-relevant problems on near term quantum devices. 
\end{abstract}

\maketitle

Combinatorial optimizations are ubiquitous in industry and technology~\cite{Korte2011}.
The potential of quantum computers for these problems has been extensively studied
~\cite{Farhi2014,Guerreschi2019,Akshay2020,Akshay2021,Wurtz2021,
Pagano2020,Harrigan2021,Ebadi2022,Yarkoni_2022}.
However, it is unclear whether they will deliver advantages in practice before fault-tolerant devices appear.
With only quadratic asymptotic runtime speed-ups expected in general~\cite{durr1996quantum, Ambainis2004,Montanaro2020} and low clock-speeds~\cite{Babbush2021, Campbell_2019}, a major challenge is the number of qubits required for quantum solvers to become competive with classical ones.
Current implementations are restricted to noisy intermediate-scale quantum devices~\cite{Preskill2018}, with variational quantum algorithms~\cite{Cerezo2021,Bharti2022} as a promising alternative. 
These are heuristic models -- based on parameterized quantum circuits -- that, although conceptually powerful, face inherent practical challenges~\cite{cerezo2023does, Stilck2021,Bittel2021,Anschuetz2022,McClean2018,Wang2021,Garcia2023}. 
Among them, hardware noise is particularly serious, since its detrimental effect rapidly grows with the number of qubits. This can flatten out the optimization landscape -- causing exponentially-small gradients (barren plateaus)~\cite{Wang2021} or underparametrization~\cite{Garcia2023} -- or render the algorithm classically simulable~\cite{Stilck2021}.
Hence, near-term quantum optimization solvers are unavoidably restricted to problem sizes that fit within a limited number of qubits.

In view of this, interesting qubit-efficient schemes have been explored \cite{fuller2021approximate,Patti2022multibasis,Tan2021,huber2023exponential,leonidas2023qubit,tenecohen2023variational,Perelshtein_2023}. 
In Refs. \cite{fuller2021approximate,Patti2022multibasis}, two or three variables are encoded into the (three-dimensional) Bloch vector of each qubit, allowing for a linear space compression. 
In contrast, the schemes of \cite{Tan2021,huber2023exponential,leonidas2023qubit,tenecohen2023variational,Perelshtein_2023} encode the $m$ variables into a quantum register of size $\mathcal{O}\big(\log(m)\big)$.
However, such exponential compressions both render the scheme classically simulable efficiently and seriously limit the expressivity of the models \cite{Tan2021,tenecohen2023variational}.
Moreover, in Refs. \cite{Tan2021,huber2023exponential,leonidas2023qubit,tenecohen2023variational,Perelshtein_2023}, binary problems are relaxed to quadratic programs. 
These simplifications strongly affect the quality of the solutions.
In addition, the measurements required by those methods can be statistically demanding.
For instance, in a deployment with $m=3964$ variables \cite{leonidas2023qubit}, most of the measurement outcomes needed did not occur and were replaced by classical random bits, leading to a low quality solution compared to state-of-the-art solvers. 
To the best of our knowledge, no experimental quantum solver has so far produced non-trivial solutions to problems with $m$ beyond a few hundreds \cite{Harrigan2021,Ebadi2022,Yarkoni_2022,abbas2023quantum}. 
Furthermore, the interplay between qubit-number compression, loss function non-linearity, trainability, and solver performance in general is mostly unknown. 

\begin{figure*}[ht!]
    \centering
    \includegraphics[width=1\textwidth]{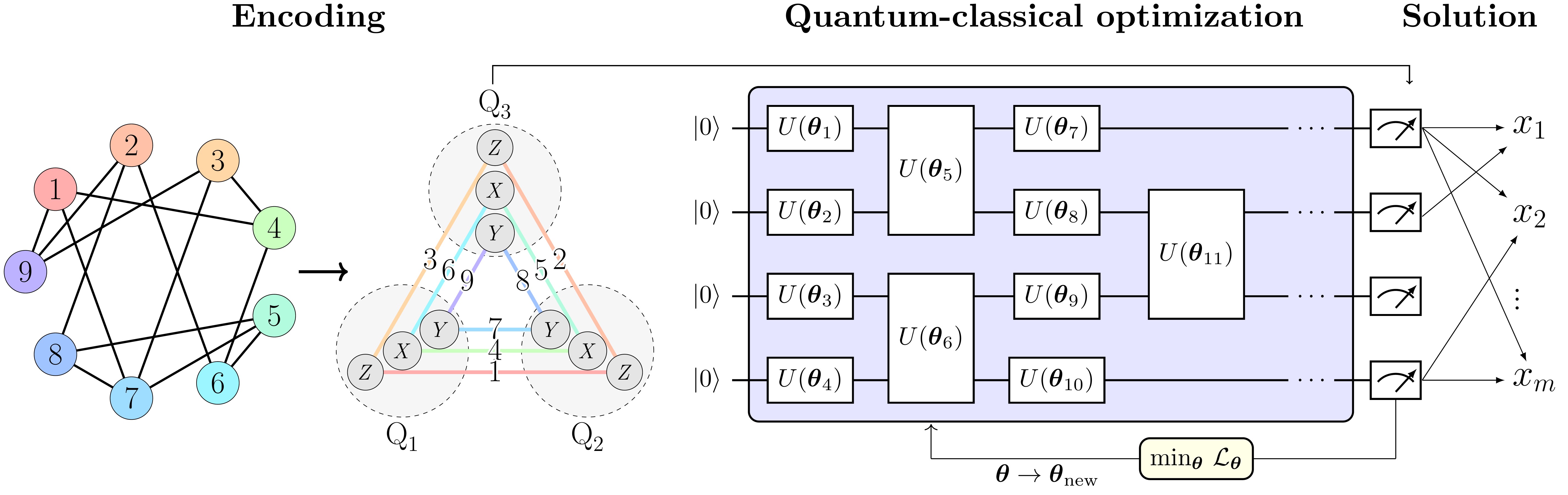}
    \caption{
    {\bf Quantum optimization solvers with polynomial space compression.} 
    Encoding: An exemplary MaxCut (or weighted MaxCut) problem of $m=9$ vertices (graph on the left) is encoded into  2-body Pauli-matrix correlations across $n=3$ qubits (Q$_1$, Q$_2$, Q$_3$). The colour code indicates which Pauli string encodes which vertex. For instance, the binary variable $x_1$ of vertex 1 is encoded in the expectation value of $Z_1\otimes Z_2\otimes\openone_3$, supported on qubits 1 and 2, while $x_9$ is encoded in $Y_1\otimes\openone_2\otimes Y_3$, over qubits 1 and 3 (see  Eq.~\eq{cut_assignment}). 
    This corresponds to a quadratic space compression of $m$ variables into $n=\mathcal{O}(m^{1/2})$ qubits.
    More generally, $k$-body correlations can be used to attain polynomial compressions of order $k$. 
    The Pauli set chosen is composed of three subsets of mutually-commuting Pauli strings. This allows one to experimentally estimate all $m$ correlations using only 3 measurement settings throughout. 
    Quantum-classical optimization: we train a quantum circuit parametrized by gate parameters $\bt$ using a loss function $\calL$ of the Pauli expectation values that mimics the MaxCut (or weighted MaxCut) objective function (see Eq. \eq{LPi}). 
    The variational Ansatz is a brickwork circuit with the number of 2-qubit gates and variational parameters scaling very mildly with $m$ (see Fig. \ref{fig:numerics}), and circuit depth sublinear in $m$. 
    This makes both experimentally- and training-friendly (see Fig. \ref{fig:barrenplateau}).
    Solution: once the circuit is trained, we read-out its output $\boldsymbol{x}$ from the correlations across single-qubit measurement outcomes on its output state. Finally, we perform an efficient classical bit-swap search around $\boldsymbol{x}$ to find potential better solutions nearby. The result of that search, $\boldsymbol{x}^*$, is the final output of our solver. 
    }
    \label{fig:main}
\end{figure*}
Here we explore this territory. We introduce a hybrid quantum-classical solver for binary optimization problems of size $m$ polynomially larger than the number of qubits $n$ used. 
This is an interesting regime in that the scheme is highly qubit-efficient while at the same time preserving the classical intractability in $m$, leaving room for potential quantum advantages.
We encode the $m$ variables into Pauli correlations across $k$ qubits, for $k$ an integer of our choice. 
A parameterized quantum circuit is trained so that its output correlations minimize a
non-linear loss function suitable for gradient descent.  
The solution bit string is then obtained via a simple classical post-processing of the measurement outcomes, which includes an efficient local bit-swap search to further enhance the solution's quality. 
Moreover, a beneficial, intrinsic by-product of our scheme is a super-polynomial suppression of the decay of gradients, from barren plateaus of heights $2^{-\Theta(m)}$ with single-qubit encodings to $2^{-\Theta(m^{1/k})}$ with Pauli-correlation encodings.  
In turn, the  circuit depth scales sublinearly in $m$, as $\mathcal{O}(m^{1/2})$ for quadratic ($k=2$) compressions and $\mathcal{O}(m^{2/3})$ for cubic ($k=3$) ones.
All these features make our scheme more experimentally- and training-friendly than previous quantum schemes,
 leading to significantly higher quality of solutions. 
 
For example, for $m=2000$ and $m=7000$ MaxCut instances, our numerical solutions are competitive with those of semi-definite program relaxations, including the powerful Burer-Monteiro algorithm. 
This is relevant as a basis for quantum-inspired classical solvers.
In addition, we deploy our solver on IonQ and Quantinuum quantum devices, observing an impressive performance even without quantum error mitigation. For example, for a MaxCut instance with $m = 2000$ vertices encoded into $n = 17$ trapped-ion qubits, we obtain estimated approximation ratios above the hardness threshold $r \approx 0.941$.
This is the highest quality reported by an experimental quantum solver on sizes beyond a few tens for MaxCut \cite{Harrigan2021,abbas2023quantum} and a few hundreds for combinatorial optimizations in general \cite{Yarkoni_2022,Ebadi2022}.
Our results open up a promising framework to develop competitive solvers for large-scale problems with small quantum devices. 
\section*{Results}
\label{sec:results}
\subsection{Quantum solvers with polynomial space compression}
\label{ssec:PSE}
We solve combinatorial optimizations over $m=\order{n^k}$ binary variables using only $n$ qubits, for $k$ a suitable integer of our choice. 
Such a compression is achieved by encoding the variables into $m$ Pauli-matrix correlations across multiple qubits. 
More precisely, with the short-hand notation $[m]:=\{1,2, \hdots m\}$, let 
$\boldsymbol{x}:=\{x_i\}_{i\in[m]}$ denote the string of optimization variables and choose a specific subset $\Pi:=\{\Pi_i\}_{i\in[m]}$ of $m\le4^n-1$ traceless Pauli strings $\Pi_i$, i.e. of $n$-fold tensor products of identity ($\openone$) or Pauli ($X$, $Y$, and $Z$) matrices, excluding the $n$-qubit identity matrix $\openone^{\otimes n}$. 
We define a {\it Pauli-correlation encoding} (PCE) relative to $\Pi$ as 
\begin{align}\label{eq:cut_assignment}
    x_i := \text{sgn}\big(\langle \Pi_i\rangle\big) \text{ for all }  i\in[m],
\end{align}
where $\text{sgn}$ is the sign function and $\langle \Pi_i\rangle:=\bra{\Psi}\Pi_i\ket{\Psi}$ is the expectation value of $\Pi_i$ over a quantum state $\ket{\Psi}$.
In \nameref{ssec:frustration} in SI, we prove that expectation values of magnitude $\Theta(1/m)$ are enough to guarantee the existence of such states for all bit strings $\boldsymbol{x}$. In practice, however, we observe magnitudes significantly larger than $\Theta(1/m))$ (see Fig. \ref{fig:histogram_distributions} in SI).
We focus on strings with $k$ single-qubit traceless Pauli matrices. 
In particular, we consider encodings $\Pi^{(k)}:=\big\{\Pi^{(k)}_1,\ldots,\Pi^{(k)}_m\big\}$ where each $\Pi^{(k)}_i$ is a permutation of either $X^{\otimes k}\otimes \openone^{\otimes n-k}$, $Y^{\otimes k}\otimes \openone^{\otimes n-k}$, or $Z^{\otimes k}\otimes \openone^{\otimes n-k}$ (see left panel of Fig.~\ref{fig:main} for an example with $k=2$).
That is, $\Pi^{(k)}$ is the union of 3 sets of mutually-commuting strings.
This is experimentally convenient, since only three measurement settings are required throughout. 
Using all possible permutations for the encoding yields $m=3\binom{n}{k}$. 
In this work, we deal mostly with $k=2$ and $k=3$, corresponding to 
$m=\frac{3}{2}n(n-1)$ and $m=\frac{1}{2}n(n-1)(n-2)$, respectively. 
The single-qubit encodings of \cite{fuller2021approximate,Patti2022multibasis}, in turn, correspond to PCEs with $k=1$. 

The specific problem we solve is weighted MaxCut, a paradigmatic NP-hard optimization problem over a weighted graph $G$, defined by a (symmetric) adjacency matrix $W\in\mathbb{R}^{m\times m}$. Each entry $W_{ij}$ contains the weight of an edge $(i,j)$ in $G$. 
The set $E$ of edges of $G$ consists of all $(i,j)$ such that $W_{ij}\neq 0$.
We denote by $|E|$ the cardinality of $E$. 
The special case where all weights are either zero or one defines the (still NP-hard) MaxCut problem, where each instance is fully specified by $E$ (see \nameref{ssec:maxcut}). 
The goal of these problems is to maximize the total weight of edges cut over all possible bipartitions of $G$. 
This is done by maximizing the quadratic objective function $\mathcal{V}(\boldsymbol{x}) := \sum_{(i,j)\in E} W_{ij}(1-\,x_i\,x_j)$ (the cut value). 

We parameterize the state in Eq. \eqref{eq:cut_assignment} as the output of a quantum circuit with parameters $\boldsymbol{\theta}$,  $\ket{\Psi}=\ket{\Psi(\boldsymbol{\theta})}$, and optimize over $\boldsymbol{\theta}$ using a variational approach \cite{Cerezo2021,Bharti2022} (see also \nameref{ssec:parent} in  SI for alternative ideas on how to optimize the state). 
As circuit Ansatz, we use the brickwork architecture shown in Fig. \ref{fig:main} (see \nameref{ssec:numdetails} for details on the variational Ansatz). 
The goal of the parameter optimization is to minimize the non-linear loss function
\begin{align}\label{eq:LPi}
    \calL &= \!\!\sum_{(i,j)\in E} \!\!\!W_{ij}\tanh\big(\alpha\,\langle \Pi_i\rangle\big)\tanh\big(\alpha\,\langle \Pi_j\rangle\big) + \calL^{(\text{reg})}.
\end{align}
The first term  corresponds to a relaxation of the binary problem where the sign functions in Eq. \eq{cut_assignment} are replaced by smooth hyperbolic tangents, better-suited for gradient-descent methods \cite{Patti2022multibasis}. 
The second term, $\calL^{(\text{reg})}$ (see \nameref{ssec:reg} in SI), forces all correlators to go towards zero, which is observed to improve the solver's performance (see \nameref{si_sec:loss_func} in the Supplementary Information). 
However, too-small correlators restrict the $\tanh$ to a linear regime ($\tanh(z)\approx z$ for 
$|z|<<1$), which is inconvenient for the training. 
Hence, to restore a non-linear 
response, we introduce a rescaling factor 
$\alpha>1$. We observe a good performance for the choice $\alpha\approx n^{\lfloor k/2 \rfloor }$ (see \nameref{ssec:hyperparameters} in SI). 

Once the training is complete, the circuit output state is measured and a bit-string $\boldsymbol{x}$ is obtained via Eq.~\eq{cut_assignment}. 
Then, as a classical post-processing step, we perform one round of single-bit swap search (of complexity $\mathcal{O}(\abs{E})$) around $\boldsymbol{x}$ in order to find potential better solutions nearby (see \nameref{ssec:numdetails}). 
The result of the search, $\x^*$, with cut value $\mathcal{V}(\boldsymbol{x}^*)$, is the final output of our solver. 

Our work differs from  \cite{Tan2021,huber2023exponential,leonidas2023qubit,tenecohen2023variational,Perelshtein_2023} in  fundamental ways.   
First of all, as mentioned, those studies focus mainly on exponential compressions in qubit number. 
These are also possible with PCEs, since there are $4^n-1$ traceless operators available.
However, besides automatically rendering the schemes classically simulable  efficiently \cite{Tan2021,tenecohen2023variational}, exponential compressions strongly limit the expressivity of the model, since $L$-depth circuits contain $\mathcal{O}(L\times \log(m))$ parameters.
This affects the quality of the solutions.
Conversely, our method operates manifestly in the regime of classically intractable quantum circuits
.  
Secondly, as for experimental feasibility, while the previous schemes require the measurement of probabilities that are (at best) of order $m^{-1}$, our solver is compatible with significantly larger expectation values 
(see Fig. \ref{fig:histogram_distributions} in SI).
Third, while in \cite{Tan2021,huber2023exponential,leonidas2023qubit,tenecohen2023variational} the problems are relaxed to quadratic programs \cite{Tan2021}, Eq. \eqref{eq:LPi} defines a highly non-linear optimization. 
These 
features lead to solutions notably superior to those of previous schemes.

\subsection{Circuit complexities and approximation ratios}
\label{ssec:numerics}
\begin{figure}[t!]
\centering
    \begin{subfigure}
        \centering
        \includegraphics[width=.98\linewidth]{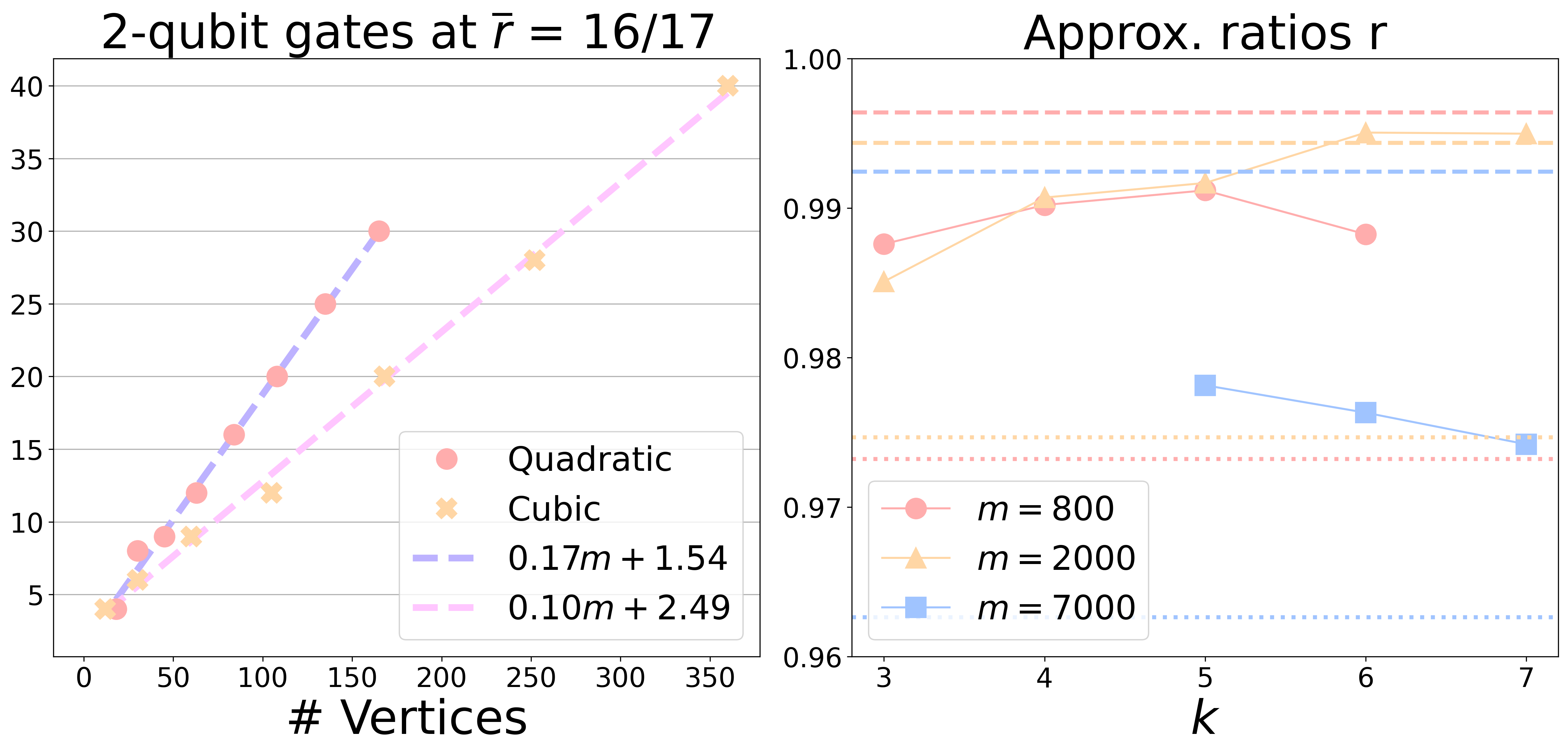}
    \end{subfigure} 
    \caption{{\bf Gate complexity and performance.} 
    Left: Number of two-qubits gates needed for achieving an average estimated approximation ratio $\overline{r}\ge 16/17\approx 0.941$ (over 250 non-trivial random MaxCut instances and 5 random initializations per instance) without the local bit-swap search (quantum-circuit's output $\boldsymbol{x}$ alone) versus $m$, both for quadratic and cubic compressions. A linear scaling is observed in both cases. 
    Right: Maximum $r$ (now including the local bit-swap search step) over random initializations for three specific MaxCut instances of different sizes as functions of the compression degree $k$ ($10$ random initializations were used for $m=800$ and $m=2000$, and $5$ for $m=7000$). 
    For a fair comparison, the total number of parameters is kept the same for all $k$. 
    The horizontal lines denote the reported results of the leading gradient-based SDP solver \cite{Choy_2000} (dotted lines) and the powerful Burer-Monteiro algorithm \cite{Burer_Monteiro_2003,Dunning2018SystematicEvaluation} (dashed lines). 
    Remarkably, our solver outperforms the former in all cases and even the latter for the $m=2000$ instance at $k=6$ and 7.}
    \label{fig:numerics}
\end{figure}

Here, we investigate the quantum resources (circuit depth, two-qubit gate count, and number of variational parameters) required by our scheme. 
Due to the strong reduction in qubit number, an increase in required circuit depth is expected to maintain the same expressivity. 
We benchmark on graph instances whose exact solution $\mathcal{V}_\text{max}:=\max_{\x} \mathcal{V}(\x)$ is unknown in general. Therefore, we denote by $r_\text{exact}:=\mathcal{V}(\x^*)/\mathcal{V}_\text{max}$ the exact approximation ratio and by $r:=\mathcal{V}(\x^*)/\mathcal{V}_\text{best}$ the estimated approximation ratio based on the best known solution $\mathcal{V}_\text{best}$ available (see \nameref{ssec:numdetails}).\\
In Fig.~\ref{fig:numerics} (left panel), we plot the gate complexity required to reach $\overline{r}= 16/17 \approx 0.941$ without doing the final local search step (to capture the resource scaling exclusively due to the quantum subroutine) on non-trivial random MaxCut instances of increasing sizes, for the encodings $\Pi^{(2)}$ and $\Pi^{(3)}$. 
For $r_\text{exact}$, this value gives the threshold for worst-case computational hardness. By non-trivial  instances we mean instances post-selected to discard easy ones (see \nameref{ssec:numdetails}). 
The results suggest that the number of gates scales approximately linearly with $m$. 
The same holds also for the number of variational parameters, which is proportional to the number of gates. 
In turn, the number of circuit layers scales as $\mathcal{O}(m/n)$. For quadratic and cubic compressions, e.g., this corresponds to $\mathcal{O}(m^{1/2})$ and $\mathcal{O}(m^{2/3})$, respectively.   
These surprisingly mild scalings translate directly into experimental feasibility and model-training ease. 
In fact, we observe (see \nameref{ssec:training_cpxt} in SI) that the number of  epochs needed for training also scales linearly with $m$. 
Moreover, in \nameref{ssec:sample} in SI, we prove worst-case upper bounds on the number of measurements required to estimate $\calL(\bt)$. For $k=2$ and $k=3$, e.g., these bounds coincide and give $\tilde{\mathcal{O}}\left(m\,(6|E|+m)^2\right)$. 

In Fig.~\ref{fig:numerics} (right), in turn, we plot solution qualities versus $k$, for three MaxCut instances from the benchmark set \verb|Gset| \cite{Gset} (see \nameref{ssec:numdetails}). 
The total number of variational parameters is fixed by $m$ (or as close to $m$ as allowed by the circuit ansatz) for a fair comparison, with the circuit depths adjusted accordingly for each $k$. 
In all cases, $r$ increases with $k$ up to a maximum, after which the performance degrades.  
This is consistent with a limit in compression capability before compromising the model's expressivity, as expected. 
Remarkably, the results indicate that our solutions are competitive with those of state-of-the-art classical solvers, such as the leading gradient-based SDP solver \cite{Choy_2000}, based on the interior points method, and even the Burer-Monteiro algorithm \cite{Burer_Monteiro_2003,Dunning2018SystematicEvaluation}, based on non-linear programming. 
Importantly, while our solver performs a single optimization followed by a single-bit swap search, the Burer-Monteiro algorithm includes multiple re-optimizations and two-bit swap searches (see \nameref{ssec:extra_numerics} in SI). 
This highlights the potential for further improvements of our scheme.
All in all, the impressive performance seen in Fig.~\ref{fig:numerics} is not only relevant for quantum solvers, but also suggests our scheme as an interesting heuristic for quantum-inspired classical solvers.
\begin{figure}[ht]
\centering
    \includegraphics[width=1\linewidth]{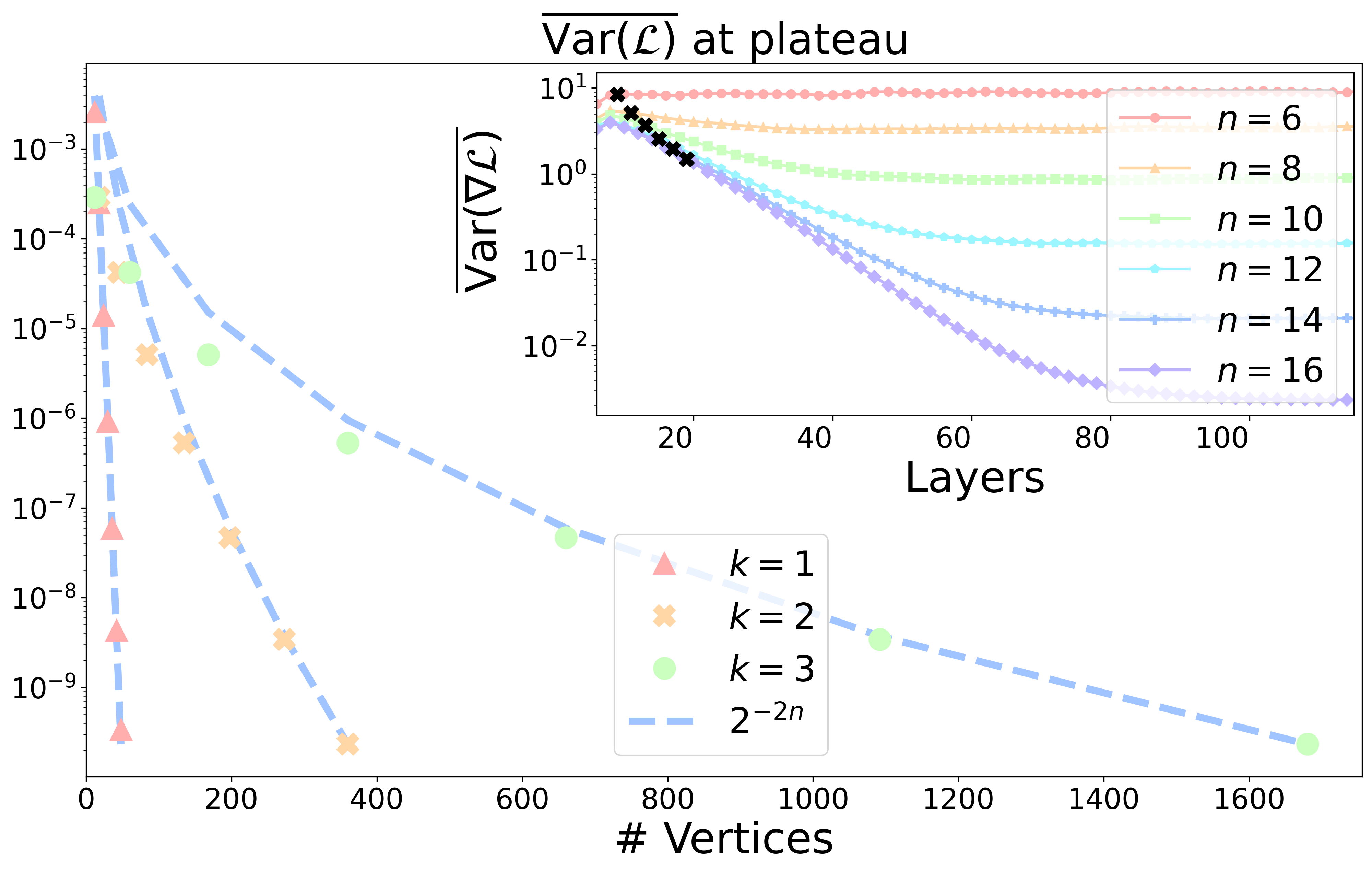}
    \caption{ {\bf Loss-function variance decay.} 
    Main: Average sample variance $\overline{\text{Var}(\calL)}$ of $\calL$, normalized by $\alpha^4 \sum_{(i.j)\in E} w_{ij}^2$, after plateauing, for the encodings $\Pi^{(1)}$, $\Pi^{(2)}$, and $\Pi^{(3)}$, as a function of $m$ (in log-linear scale). 
    The agreement with the analytical expression in Eq. \eqref{eq:variance} is excellent (the dashed, blue curve corresponds to the first term of the equation, which decreases as $2^{-2n}$).
    Since $n=\mathcal{O}(m^{1/k})$, this translates into a super-polynomial suppression in $m$ of the decay speed of $\overline{\text{Var}(\calL)}$ for $k>1$. 
    Inset: average variances of the entries of the gradient of $\calL$ as functions of the number of layers, for quadratic compression ($\Pi^{(2)}$). Each curve corresponds to a different $n$. 
    Note the decay of the plateau (rightmost) values with $n$.  
    The black crosses indicate the depths needed to reach average approximation ratios $>0.941$ (computational-hardness threshold) with the quantum-circuit's output $\boldsymbol{x}$ alone, i.e. excluding the final bit-swap search. 
    Remarkably, in all cases such ratios are attained 
    before the variances have converged to their asymptotic, steady values. 
    }
    \label{fig:barrenplateau}
\end{figure}

\subsection{Intrinsic mitigation of barren plateaus}
\label{ssec:bp}

Another appealing feature of our solver emerging from the qubit-number reduction is an intrinsic mitigation of the infamous barren plateau (BP) problem \cite{McClean2018, cerezo2021cost, Wang_2021, fontana2023adjoint,ragone2023unified}, which constitutes one of the main challenges for training variational quantum algorithms. 
BPs are characterized by a vanishing expectation value of $\nabla\calL$ over random parameter initializations and an exponential decay (in $n$) of its variance. 
This jeopardizes the applicability of variational quantum algorithms in general~\cite{cerezo2023does}.
For instance, the gradient variances of a two-body Pauli correlator  on the output of universal 1D brickwork circuits are known to plateau at levels exponentially small in $n$ for circuit depths of about $10\times n$ \cite{McClean2018}. 
Alternatively, BPs can equivalently be defined in terms of an exponentially vanishing variance of $\LC$ itself (instead of its gradient)~\cite{arrasmith2022equivalence}. This is often more convenient for analytical manipulations.

In \nameref{ssec:bp_analytics} in the Supplementary Information we prove that, if the random parameter initializations make the circuits sufficiently random (namely, induce a Haar measure  over the special unitary group), the variance of $\LC$ is given by 
\begin{equation} \label{eq:variance}
    {\rm Var}(\LC) = \frac{\alpha^4}{d^2} \sum_{(i,j)\in E} w^2_{ij} + \OC\left(\frac{\alpha^6}{d^3}\right)\,,
\end{equation}
where $d=2^n$ is the Hilbert-space dimension. 
Interestingly, the leading term in Eq. \eqref{eq:variance} appears also if one only assumes the circuits to form a $4$-design, but it is then not clear how to bound the higher-order terms without the full Haar-randomness assumption. However, we suspect that the latter is indeed not necessary.
In practice, for 1D brick-work random quantum circuits, the unitary-design assumption is approximately met at depth $\mathcal{O}(n)$ \cite{BHH16,Haferkamp_2022}. 
In line with that, for our loss function, we empirically observe convergence to Eq.~\eqref{eq:variance} at circuit depths of about $8.5\times n$. This is illustrated in Fig.~\ref{fig:barrenplateau} for linear, quadratic, and cubic compressions, where we plot the average sample variance $\overline{\text{Var}(\calL)}$ of $\calL$ over 100 non-trivial random MaxCut instances and 100 random parameter initializations per instance, as a function of $m$. In contrast, the depth needed to reach $r>0.941$ on average with the circuit's output alone is about $ 1.05 \times n$ (see figure inset).

One observes an excellent agreement between $\overline{\text{Var}(\calL)}$ and the first term of Eq. \eqref{eq:variance} for large $m$.
As $m$ decreases, small discrepancies appear, specially for $k=2$ and $k=3$. 
This can be explained by noting that $\alpha\sim 1.5$ for $k=1$ whereas $\alpha\sim 1.5\times n$ for $k=2$ and $k=3$ (see \nameref{ssec:hyperparameters} in SI), so that the second term in \eqref{eq:variance} scales as $2^{-3n}$ for the former but as $n^{6}\,2^{-3n}$ for the latter. 
Hence, as $m$ (and so $n$) decreases, that term requires smaller $m$ to become non-negligible for the former than for the latter.
Remarkably, the scaling $\overline{\text{Var}(\calL)}\in\Theta(\alpha^4 \,2^{-2n})$ in $n$ translates into a super-polynomial suppression of the decay speed in $m$ when compared to single-qubit (linear) encodings. 
This means, for instance, that quadratic encodings feature $\overline{\text{Var}(\calL)}\in\Theta(\alpha^4 \,2^{-2\sqrt{m}})$, instead of $\overline{\text{Var}(\calL)}\in\Theta(\alpha^4 \,2^{-2\,m})$ displayed by linear encodings. 
Importantly, the scaling obtained still represents a super-polynomial decay in $m$. Yet, the enhancement obtained makes a tremendous difference in practice, as shown in the figure by the orders of magnitude separating the three curves.

\begin{figure}[t]
\centering
    \includegraphics[width=1
\linewidth]{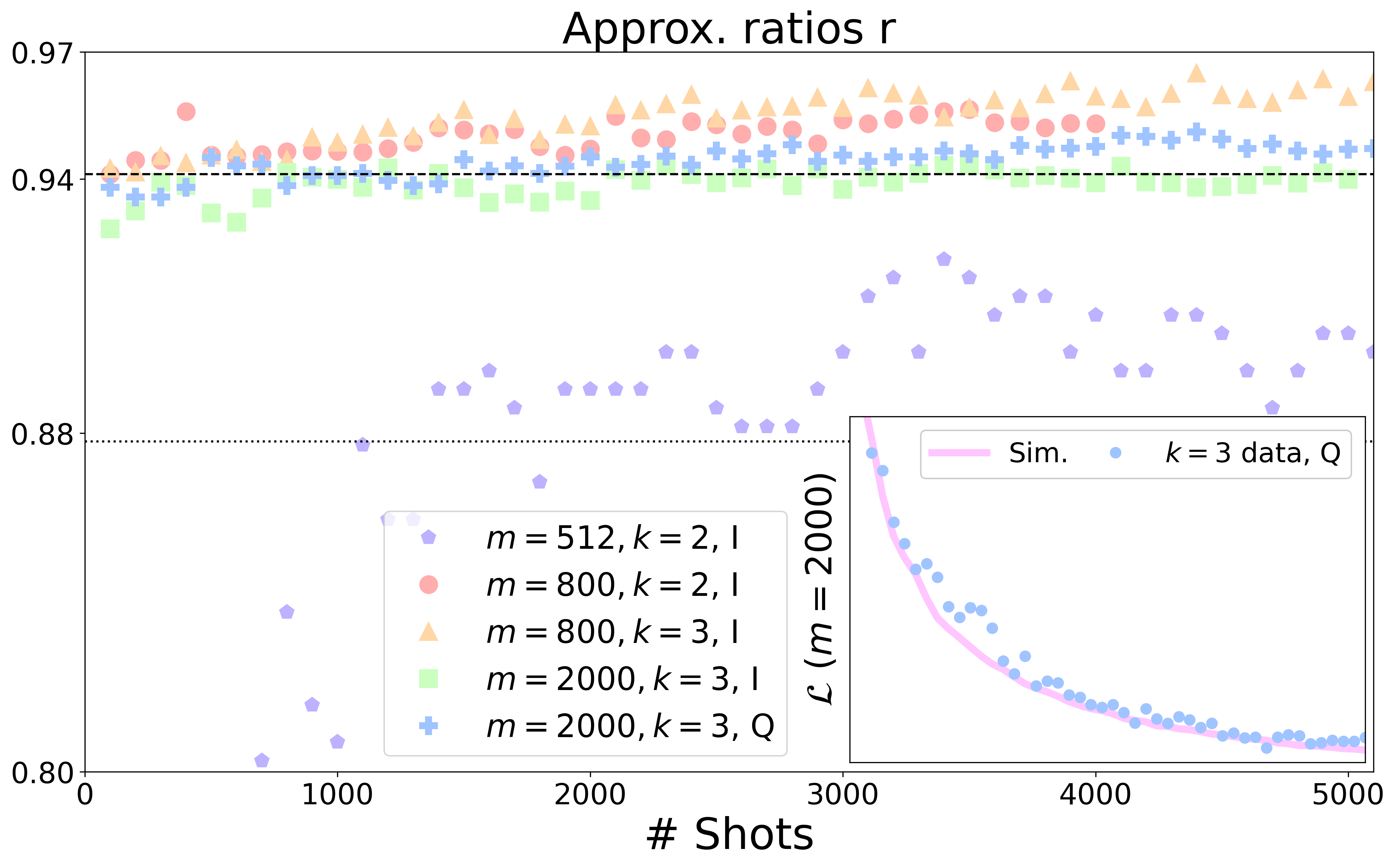}    
    \caption{{\bf Trapped-ion experimental implementation}. Main:
    estimated approximation ratios for our scheme deployed on IonQ's Aria-1 (I) and Quantinnum H1-1 (Q) devices as functions of the number of measurements (per each of the three measurement settings). Three problem instances (see main text for details) are shown: one weighted MaxCut instance of $m=512$, solved with quadratic compression using $n=19$ qubits (purple pentagons), one MaxCut instance of $m=800$, solved with quadratic compression on $n=23$ qubits (red circles) and cubic compression on $n=13$ qubits (yellow triangles), and another MaxCut instance of $m=2000$, solved with  cubic compression on $n=17$ qubits (blue squares).  
    The black horizontal lines indicate the Goemans-Williamson threshold  (dotted), at $r\approx0.878$, and the worst-case computational-hardness threshold (dashed), at $r=16/17$. 
    Inset:  
    loss function  $\calL$ (at fixed, optimized parameters) versus number of measurement shots 
(same shot range as in main figure), for the $m=2000$ instance with cubic compression. The solid, pink curve corresponds to our numerical simulation, while the blue dots are the experimental data for the implementation on Quantinuum (the highest-fidelity one)
.
    }
    \label{fig:results_experiment}
\end{figure} 
\subsection{Experimental deployment on quantum hardware}
We experimentally demonstrate our quantum solver on IonQ's Aria-1 and Quantinuum H1-1 trapped-ion devices, for two MaxCut instances of $m=800$ and $2000$ vertices and a weighted MaxCut instance of $m=512$ vertices.
Details on the hardware and model training are provided in \nameref{ssec:expdetails}, while the choice of instances is detailed in \nameref{ssec:numdetails} (see Table \ref{tab:instances}). 
We optimize the circuit parameters offline via classical simulations and experimentally deploy the pre-trained circuit.
Fig.~\ref{fig:results_experiment} depicts the obtained approximation ratios for each instance as a function of the number of measurements, employing both the quadratic and cubic Pauli-correlation encodings, $\Pi^{(2)}$ and $\Pi^{(3)}$, respectively. 
For each instance, we collected enough statistics for the approximation ratio to converge (see figure inset). 
The circuit size is limited by gate infidelities. For IonQ, we found a good trade-off between expressivity and total infidelity at $90$ two-qubit gates altogether. 
Quantinuum's device, which displays significantly higher fidelities, allows for larger circuits, but we used the same number of gates for simplicity.
This is below the number required for these instance sizes according to 
Fig.~\ref{fig:numerics} (left), especially for $k=2$. 
However, remarkably, our solver still returns solutions of higher quality than the G\"oemans-Williamson bound in all cases and even than the worst-case hardness threshold in 
four out of the 
five experiments. 
This is the first-ever quantum experiment to produce such high-quality solutions for these sizes. 
As a reference, the largest MaxCut instance experimentally solved with QAOA~\cite{Farhi2014} has size $m=414$ and average and maximal approximation ratios 0.57 and 0.69, respectively (see Table IV in Ref. \cite{abbas2023quantum}).

\section*{Conclusions and Discussion}
\label{sec:conclusions}

We introduced a scheme for solving binary optimizations of size $m$ polynomially larger than the number of qubits used. 
Pauli correlations across few qubits encode each binary variable.
The circuit depth is sublinear in $m$, while the numbers of parameters and training epochs approximately linear in $m$. Moreover, the qubit-number compression brings in the 
beneficial by-product of significantly suppressing the decay in $m$ of the variances of the loss function (and its gradient), which we have both analytically proven and verified numerically. These features, together with an educated choice of non-linear loss function, allow us to solve large, computationally non-trivial instances with unprecedentedly-high quality. 
Numerically, our solutions for $m=2000$ and $m=7000$ MaxCut instances are competitive with those of 
state-of-the-art solvers such as the powerful Burer-Monteiro algorithm.
Experimentally, in turn, for a deployment on $17$ trapped-ion qubits, we estimated approximation ratios beyond the worst-case computational hardness threshold $0.941$ for a non-trival MaxCut instance with $m=2000$ vertices.  
To our knowledge, this the highest solution quality ever reported experimentally on such instance sizes. 

We stress that these results are based on raw experimental data, without any quantum error mitigation procedure to the observables measured. 
Yet, our method is indeed well-suited for standard error mitigation techniques \cite{PhysRevX.7.021050,PhysRevLett.119.180509,cai2023quantum}, the use of which can enhance the solver's performance even further.
In turn, although we have focused on quadratic unconstrained binary optimization (QUBO) problems, the technique can be straightforwardly extended to generic polynomial unconstrained binary optimizations (PUBOs) \cite{chermoshentsev2022polynomial} without any increase in qubit numbers. This is in contrast to conventional PUBO-to-QUBO reformulations, which incur in expensive overheads in extra qubits \cite{Biamonte_2008}. 
Interestingly, for certain problems with specific structure, such as for instance the traveling salesperson problem, PUBO reformulations exist that are more qubit-efficient than the corresponding QUBO versions \cite{Glos22}. Combining such reformulations with our techniques could allow for polynomial qubit-number reductions on top of that.

Importantly, as with most variational quantum algorithms (VQAs), an open question is the run-time of experimentally training the model. 
Our loss function's gradients can be estimated via the gradient chain rule together with the standard parameter shift rule \cite{Cerezo2021,Bharti2022}. 
Particularly challenging is the number of measurements required for estimating the loss function. 
If $|E|$ is linear in $m$, e.g., our analysis gives a worst-case upper bound $\tilde{\mathcal{O}}(m^3)$ to the sample complexity of estimating the loss function. 
However, we note that this is significantly better than in VQAs for chemistry, where the sample complexity of estimating the loss function (the energy) scales as the problem size (number of orbitals) to the eighth power for popular basis sets such as STO-nG \cite{RevModPhys.92.015003}. In addition,
further improvement to our sample complexity is possible by optimization of  hyperparameters ($\alpha$ and $\beta$, e.g.) on a case-by-case basis. 
Moreover, the perspectives improve even more if suitable pre-training strategies are introduced. For example, pre-training with classical tensor-network simulations can drastically reduce both circuit depth and training run-time \cite{Rudolph23}.
Another potentially relevant tool for pre-training is given in \nameref{ssec:parent} in SI, where we derive Hamiltonians whose ground states give approximate MaxCut solutions via our multi-qubit encoding
. Such Hamiltonians may be used for QAOA schemes \cite{Farhi2014} to prepare warm-start input states for the core variational circuit.

Finally, other exciting open questions are the role of entanglement in our solver and the relation between our method and purely classical schemes where, instead of a quantum circuit, generative models are used to produce the correlations (see \nameref{ssec:time_complexity} in SI). 
However, as for the latter, the fact that our circuits cannot be classically simulated efficiently gives  our approach interesting prospects. 
All in all, our framework offers a promising machine-learning playground to explore 
quantum optimization solvers on large-scale problems, both with small quantum devices in the near term and with quantum-inspired classical techniques.

\section*{Methods}
\label{sec:methods}

\subsection{MaxCut problems}
\label{ssec:maxcut}

The weighted MaxCut problem is an ubiquitous
combinatorial optimization problem. 
It is a graph partitioning problem defined on weighted undirected graphs $G=(V,E)$ whose goal is to divide the $m$ vertices in $V$ into two disjoint subsets in a way that maximizes the sum of edge weights $W_{ij}$ shared by the two subsets -- the so-called cut value. 
If the graph $G$ is unweighted, that is, if $W_{ij}=1$ or  $W_{ij}=0$ for every edge $(i,j)\in E$, the problem is referred to simply as MaxCut. 
By assigning a binary label $x_i$ to each vertex $i\in V$, 
the problem can be mathematically formulated as the binary optimization
\begin{equation}
    \underset{\boldsymbol{x}\in \{-1, 1\}^{m}}{\text{maximize}} \hspace{0.2cm} \sum_{i,j\in[m]} W_{ij}(1-\,x_i\,x_j) \,.   
    \label{eq:maxcut}
\end{equation}
Since $\sum_{i,j\in[m]} W_{ij}$ is constant over $\boldsymbol{x}$, Eq. \eqref{eq:maxcut} can be rephrased as a minimization of the objective function $\x^\text{T} W\x$. This specific format is known as a quadratic unconstrained binary optimization (QUBO).  
For generic graphs, solving MaxCut exactly is NP-hard \cite{Karp1972}. 
Moreover, even approximating the maximum cut to a ratio $r_\text{exact}>\frac{16}{17}\approx0.941$ is NP-hard \cite{Hastad2001,Trevisan2000}. 
In turn, the best-known polynomial-time approximation scheme is the Goemans-Williamson (GW) algorithm \cite{GoemansWilliamson}, with a worst-case ratio $r_\text{exact}\approx 0.878$.
Under the Unique Games Conjecture, this is the optimal achievable by an efficient classical algorithm with worst-case performance guarantees. 
If, however, one does not require performance guarantees, there exist powerful heuristics that in practice produce cut values often higher than those of the GW algorithm. Two  examples are discussed in {\it Best solutions known} in \nameref{ssec:numdetails}.

\subsection{Regularization term}
\label{ssec:reg}
The regularization term in Eq. \eq{LPi}
penalizes large correlator values, thereby forcing the optimizer to remain in the correlator domain where all possible bit string solutions are expressible.
Its explicit form is
\begin{align}\label{eq:Lreg}
    \calL^{(\text{reg})} = \beta\,\nu \left[\frac{1}{m}\sum_{i\in V} \tanh\big(\alpha\,\langle\Pi_i\rangle\big)^2\right]^2.
\end{align}

The factor $1/m$ normalizes the term in square brackets to $\order{1}$. 
The parameter $\nu$ is an estimate of the maximum cut value: it sets the overall scale of $\calL^{(\text{reg})}$ so that it becomes comparable to the first term in Eq. \eq{LPi}. For weighted MaxCut, we use the Poljak-Turz\'ik lower bound $\nu = w(G)/2 + w(T_{\text{min}})/4$ \cite{POLJAK198699}, where $w(G)$ and $w(T_{\text{min}})$ are the weights of the graph and of its minimum spanning tree, respectively. For MaxCut, this reduces to the Edwards-Erd\"os bound \cite{sztaki2217} $\nu=|E|/2+(m-1)/4$.
Finally, $\beta$ is a free hyperparameter of the model, which we optimize over random graphs to get $\beta=1/2$. Such optimizations systematically show increased approximation ratios due to the presence of $\calL^{(\text{reg})}$ in Eq. \eq{LPi} (see \nameref{si_sec:loss_func} in SI).

\subsection{Numerical details}
\label{ssec:numdetails}
{\it Choice of instances.} The numerical simulations of Figs. \ref{fig:numerics} (left) and Fig. \ref{fig:barrenplateau} were performed on random MaxCut instances generated with the well-known \verb|rudy| graph-generator \cite{rudy} post-selected so as to filter out easy instances. The post-selection consisted in discarding graphs with less than 3 edges per node on average or those for which a random cut gives an approximation ratio $r>0.82$. The latter is sufficiently far from the Goemans-Williamson ratio $0.878$ while still allowing efficient generation.
For the numerics in Fig.~\ref{fig:numerics} (right) and the experimental deployment in Fig.~\ref{fig:results_experiment} we used 6 graphs from standard benchmarking sets: the former used the {\it G14}, {\it G23}, and {\it G60} MaxCut instances from the \verb|Gset| repository \cite{Gset}, while the latter used {\it G1} and {\it G35} from \verb|Gset| and the weighted MaxCut instance {\it pm3-8-50} from the \verb|DIMACS| library \cite{DIMACS} (recently employed also in \cite{Patti2022multibasis}). 
Their features are summarized in Table \ref{tab:instances}. \\

{\it Best solutions known.} 
For the generated instances, the best solution is taken as the one with the highest cut value between the (often coinciding) solutions produced by two classical heuristics, namely the Burer-Monteiro \cite{Burer_Monteiro_2003} and the Breakout Local Search \cite{BENLIC20131162} algorithms. For the instances from benchmarking sets, we considered instead the best known documented solution. 
The corresponding cut value, $\mathcal{V}_\text{best}$, is used to define the approximation ratio achieved by the quantum solution $\x^*$, namely $r=\mathcal{V}(\x^*)/\mathcal{V}_\text{best}$. \\

\begin{table}[h]
\begin{ruledtabular}
\begin{tabular}{c|ccccc}
{\it Graph}                     & $m$               & $|E|$            & $W_{ij}$           & {\it Type}    &  {\it Use}
\\ \hline
{\it pm3-8-50}            & 512               & 1536           & $\pm1$             & $3D$ torus grid &  Experiment    \\ 
{\it G1}                  & 800               & 19176          & $1$                & random  & Experiment  \\ 
{\it G14}                 & 800             & 4694          & $1$                & planar     & Numerics  \\ 
{\it G23}                 & 2000              & 19990          & $1$                & random   & Numerics   
\\ 
{\it G35}                 & 2000              & 11778          & $1$                & planar   & Experiment  \\ 
{\it G60}                 & 7000              & 17148          & $1$                & random  & Numerics  \ \ 

\end{tabular}
\end{ruledtabular}
\caption{{\bf Benchmark 
instances used in this work}. 
Apart from the the number of vertices, edges, and edge weights, we also include the type of 
graph as well as its use.}  \label{tab:instances}
\end{table} 

{\it Variational Ansatz.}
As circuit Ansatz, we used the brickwork architecture shown in Fig. \ref{fig:main}, with 
layers of single-qubit rotations, parameterized by a single angle, followed by a layer of M\o lmer-S\o rensen (MS) two-qubit gates, each with three variational parameters. Each single-qubit gate layer contains rotations around a single direction (X, or Y, or Z), one at a time, sequentially. Furthermore, we observed that many of the other commonly used parameterized gate displays the same numerical scalings up to a constant.  \\

{\it Quantum-circuit simulations.} 
The classical simulations of quantum circuits have been done using two libraries: \verb|Qibo| \cite{qibo_paper,qibojit_paper} for exact state-vector simulations of systems up to 23 qubits, and \verb|Tensorly-Quantum| \cite{patti2021tensorlyquantum} for tensor-network simulations of larger qubit systems. \\
 
{\it Optimization of circuit parameters.}  
Two optimizers were used for the model training. 
SLSQP from the \verb|scipy|
library was used for systems small enough to calculate the gradient using finite differences. 
In all other cases we used Adam from the \verb|torch|/\verb|tensorflow|
libraries, leveraging automatic differentiation to speed up computational time. As a stopping criterion for Adam, we halted the training after $50$ steps whose cumulative improvement to the loss function was less then $0.01$. For both optimizers, the default optimization parameters were used. \\ 

{\it Classical bit-swap search as post-processing step.} 
As mentioned, a single round of local bit-swap search is performed on the bit string $\boldsymbol{x}$ output by the trained quantum circuit. This consists of sequentially swapping each bit of $\boldsymbol{x}$ and computing the cut value of the resulting bit string. 
If the cut value improves, we retain the change. Else, the local bit flip is reverted. 
There are altogether $\Theta(m)$ local bit flips.
A bit flip on vertex $i$ affects $\Theta(d(i))$ edges, with $d(i)$ the degree of the vertex. Hence, an update of only $\Theta(d(i))$ terms in $\mathcal{V}(\boldsymbol{x})$ is required per bit flip.
The total complexity of the entire round is thus $\Theta(\abs{E})$. \\

\subsection{Experimental details}
\label{ssec:expdetails}
\begin{table}[h]
\begin{ruledtabular}
\begin{tabular}{c|ccccc|cc} 
&   &   &  & &  & \qquad\qquad $r$ & 
\\\cline{7-8}
 {\it Graph} & $k$   & $n$  & {\it 1-q} & {\it 2-q}  & {\it Epochs} &  {\it Sim.} & {\it Exp.}  
 \\  \hline
   {\it pm3-8-50}  & 2   &19 & 199  & 90  & 13485 & 0.967 & 0.921 
   \\  
 {\it G1}  & 2   &24  & 192 & 36 & 4027 & 0.954 & 0.957 
 \\  
  {\it G1}  & 3   &13  & 170 & 36   & 2022 & 0.940 & 0.965  
  \\ 
 {\it G35}  & 3  &17 & 193  &88  & 4100 & 0.935 & 0.951 
 \\ 
\end{tabular}
\end{ruledtabular}
\caption{{\bf Details about the experimentally deployed instances}.
For each instance, we display $k$, $n$, the 1-qubit and 2-qubit gate counts, and number of optimization epochs used during classical training.
The last two columns report the approximation ratios given by the classical simulation of the noiseless circuit (sim.) and the best one observed in the experiment (exp.).
We note that all ratios 
lie more than $3$ standard deviations away from the average solution obtained 
via a single-bit  search over a randomly picked bit string (see \nameref{ssec:random} in  SI).} \label{tab:resources}
\end{table} 

{\it Hardware details. } 
The experiments were deployed on IonQ's Aria-1 25-qubit device and on Quantinuum's  H1-1 20-qubit device. Both devices are based on trapped ytterbium ions and support all-to-all connectivity. 
The VQA architecture of Fig. \ref{fig:main} was adapted accordingly to hardware native gates. We used alternating layers of  partially entangling M\o lmer-S\o rensen (MS) gates and, depending on the experiment, rotation layers composed of one or two native single-qubit rotations GPI and GPI2 (see Table~\ref{tab:resources}). Since the $z$ rotation is done virtually on the IonQ Aria chip, parameterized RZ rotation were also added at the end of every rotation layer without any extra gate cost.

The native gates in Quantinuum's H1-1 chip 
are the double parameterized $\text{U}_{1q}$ gate, a virtual $z$ rotation, and the entangling arbitrary-angle two-qubit rotation RZZ.  In our experiment, the circuit pre-trained for the $m=2000$ vertices instance using IonQ native gates was transpiled into a Quantinuum native gates circuit with the same number of 1 and 2-qubit gates.    

{\it Resource analysis.} We run a total of four experimental deployments. The three selected instances were trained using exact classical simulation with Adam optimizer, as detailed in  \nameref{ssec:numdetails}. 
In an attempt to get the best possible solution within the limited depth constraints of the hardware, the stopping criteria was relaxed to allow $150$ non-improving steps. 
This resulted in a total number of training epochs considerably larger then the average case scenario (see \nameref{ssec:extra_numerics} in SI).
Table \ref{tab:resources} reports the precise quantum (number of qubits and gate count) and classical (number of epochs) resources, as well as the observed results. \\
\section*{Acknowledgments}
\label{sec:acknowledgments}

The authors would like to thank Marco Cerezo and Tobias Haug for helpful conversations. D.G.M. is supported by Laboratory Directed Research and Development (LDRD) program of LANL under project numbers 20230527ECR and 20230049DR. 
At CalTech, A.A. is supported in part by the Bren endowed chair and Schmidt Sciences AI2050 senior fellow program. 

\bibliography{arxiv_V2}

\clearpage

\appendix

\section*{Supplementary Information}
\label{sec:SI}

\begin{figure}[ht]
\centering
    \includegraphics[width=1
\linewidth]{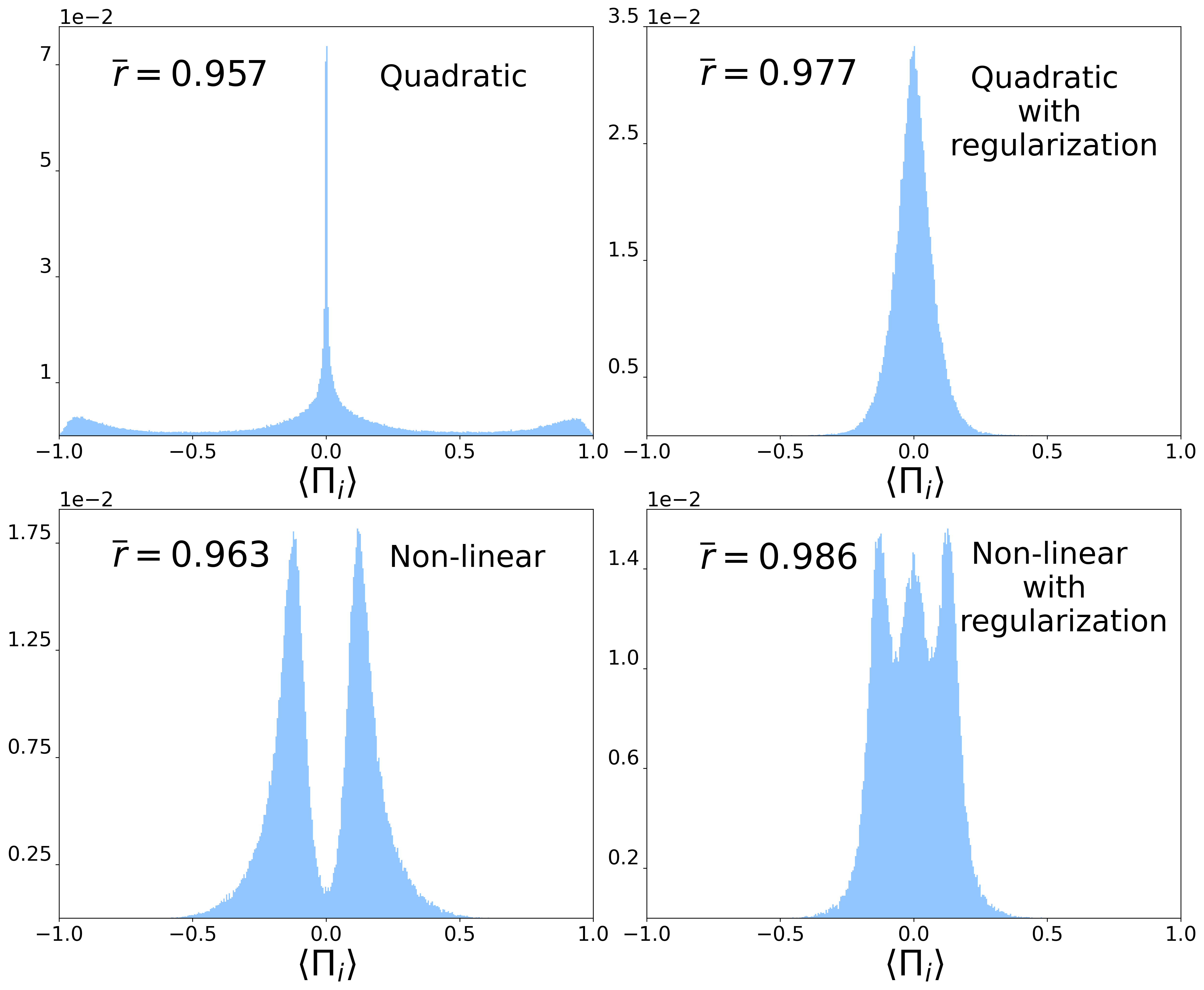}  
    \caption{{\bf Pauli correlations under different loss functions}. 
    Histograms of expectation values $\langle\Pi_i\rangle$ for the quadratic Pauli-correlation encoding $\Pi^{(2)}$ after training under different loss functions, for 250 random graph instances of $m=108$ vertices and 5 random parameter initializations per instance.
    Top left: quadratic loss function $\calL^{(\text{qua})}= \sum_{(i,j)\in E} W_{ij}\,\langle \Pi_i\rangle\,\langle \Pi_j\rangle$.
    Top right: quadratic loss function with regularization, $\calL^{(\text{qua})}+\calL^{(\text{reg})}$.
    Bottom left: non-linear loss function from Eq. \eq{LPi} with regularization removed, $\calL-\calL^{(\text{reg})}$.
    Bottom right: Complete loss function $\calL$ given by Eq. \eq{LPi}. 
    The panels show also the average approximation ratios $\overline{r}$ obtained in each case.}
    \label{fig:histogram_distributions}
\end{figure}

\section*{Choice of loss function}
\label{si_sec:loss_func}

Here we motivate the specific loss function chosen in Eq. \eq{LPi}. 
$\calL$ leverages two main features: the non-linearities from the hyperbolic tangents and the regularization term $\calL^{\text{(reg)}}$, given by \eq{Lreg}, which forces all the correlators to have small values.
In Fig. \ref{fig:histogram_distributions}, we show a comparison of the distribution of expectation values $\langle\Pi_i\rangle$ of Pauli string correlators at the end of a training process based on four different loss functions. 
Namely, we compare Eq. \eq{LPi} with similar loss functions obtained by removing the hyperbolic tangent factors (i.e., a quadratic function of the $\langle\Pi_i\rangle$) and/or the regularization term $\calL^{(\text{reg})}$. 

\begin{figure}[t]
\centering
    \includegraphics[width=1
\linewidth]{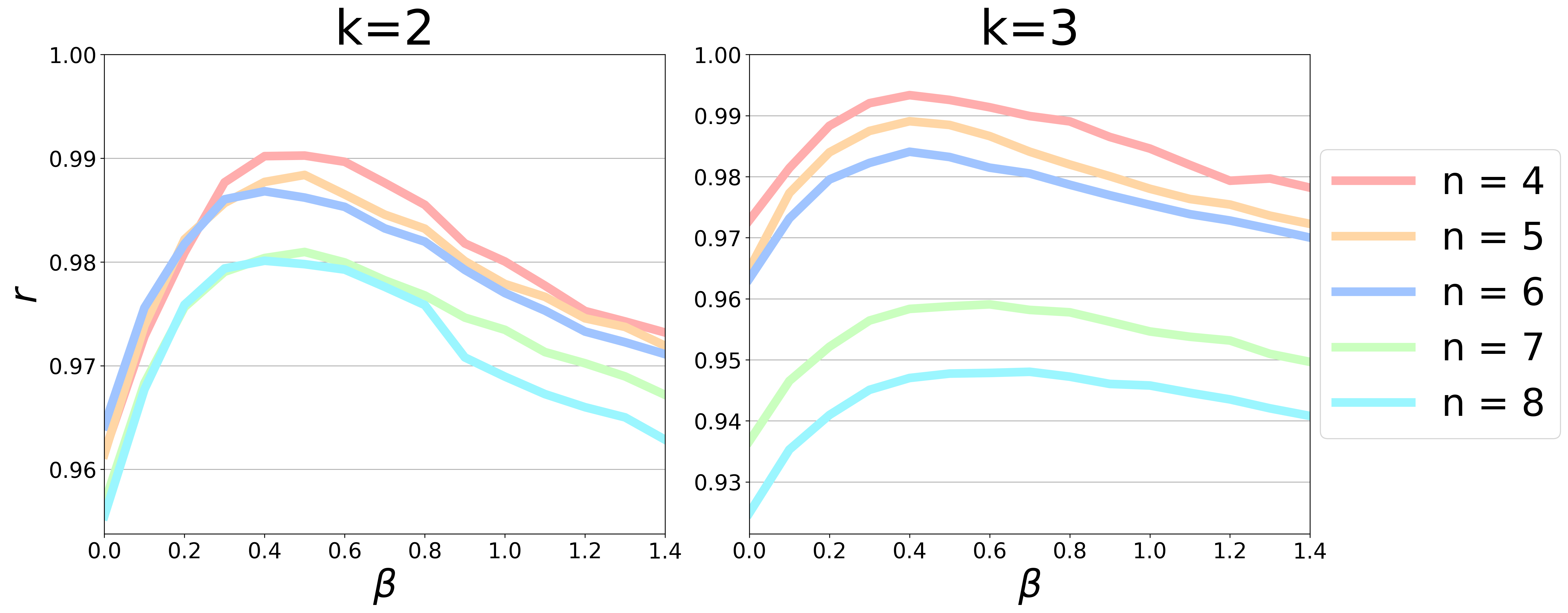}  
    \caption{{\bf Tuning of $\beta$.} The plots shows the average approximation ratio versus $\beta$ over 250 randomly generated graph instances (over 5 initialization) for increasing number of qubits, at fixed depth $l=11$ . Each plot display an order of polynomial compression: $k=2$ and $k=3$.}
    \label{fig:tuning_beta}
\end{figure}

We see that, for a quadratic loss function without the regularization term 
(top left), the distribution of expectation values is approximately flat with a peak around the origin and small peaks around $\pm 1$. 
The introduction of the non-linear function $\tanh(\cdot)$ 
(bottom left) causes the expectation values to cluster in heavy-tailed distributions around two symmetric points. 
This alters the optimization landscape, thereby discouraging extremal values, which is a particularly important feature for our encoding due to its sensitivity to frustration constraints. 
We emphasize that the specific choice of the hyperbolic tangents is not particularly important: we observed that any sigmoid-like non-linear function leads to the same concentration phenomenon; this is a consequence of their vanishing gradients close to the extrema.  
The addition of the regularization term to each of those loss functions (top and bottom right plots, respectively) further incentivizes the $\langle\Pi_i\rangle$ to stay close to zero, reducing the tails of the distributions and shifting their mean magnitude closer to zero. 
The strength of this shift is modulated by the hyperparameter $\beta$ appearing in Eq. \eq{Lreg}, whose value we also fine-tune by extensive numerical exploration on random graph instances following the same procedure of \nameref{ssec:hyperparameters} for $\alpha$. The result is shown in Fig.~\ref{fig:tuning_beta}.

\section*{Sufficient conditions for the encoding}
\label{ssec:frustration}
\begin{figure*}[ht]
    \centering
         \includegraphics[width=1\linewidth]{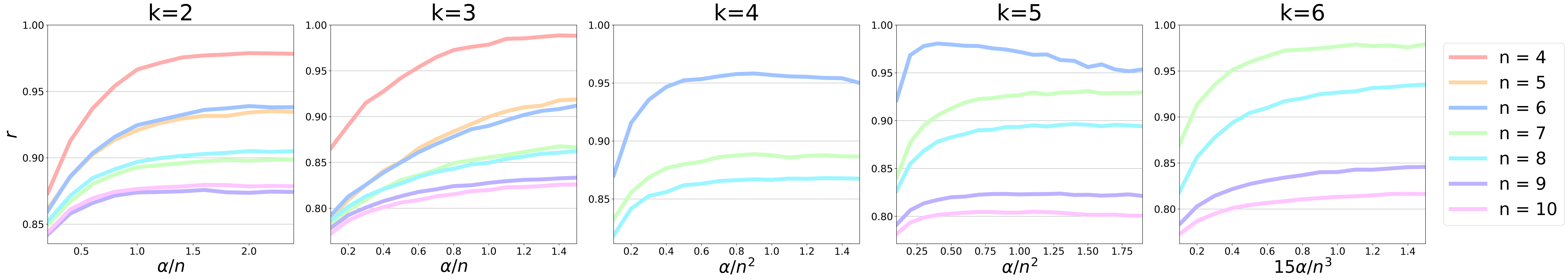}
    \caption{{\bf Tuning of $\alpha$.} Average approximation ratios versus $\alpha$ over 250 randomly generated graph instances (with 5 initializations each) for increasing number of qubits, at fixed depth $l=5$ . Each plot displays an order of polynomial compression, from $k=2$ (leftmost) to $k=6$ (rightmost). The precise value of $\alpha$ used by our solver is fine-tuned to maximize $r$. For $k=1$ (not shown), we observe a constant behaviour $\alpha \sim 1.5$. The optimal values observed are thus compatible with the scaling
    $\alpha = \mathcal{O}(n^{\lfloor k/2 \rfloor })$.
    }
    \label{fig:tuning_alpha}
\end{figure*} 

Here we derive sufficient (but not necessary) conditions on the magnitudes of Pauli string correlators for encoding arbitrary bit strings into valid quantum states as per Eq. \eqref{eq:cut_assignment}.   


For an arbitrary bit string $\boldsymbol{x}$, we define  
\begin{equation}
    \varrho_{i} = \frac{\openone + x_{i}\Pi_i}{2^n},
\end{equation}
where $x_i$ is the $i$-th bit of $\boldsymbol{x}$. The above state is clearly hermitian, trace-1, and such that $\text{Tr}[\varrho_{i}\Pi_{j}] = x_j \delta_{i, j}$. Moreover, $\varrho_{i}$ is diagonal in the eigenbasis of $\Pi_{i}$, with eigenvalues $(1 \pm x_{i})/2^{n} \geq 0$. Hence, $\varrho_{i}$ is positive semi-definite and, so, a valid density matrix. Next, we define 
\begin{equation}
    \varrho = \frac{1}{m}\sum_{i = 1}^{m} \varrho_{i} = \frac{\openone}{2^n} + \frac{1}{2^n}\sum_{i = 1}^{m}\frac{x_{i}}{m}\Pi_{i}.
\end{equation}
Since this is a convex combination of positive semi-definite matrices, it it also positive semi-definite. Moreover it satisfies
\begin{equation}\label{eq:mixed_state}
    \text{Tr}[\varrho \, \Pi_{i}] = \frac{x_{i}}{m},
\end{equation}
for all $i \in [m]$. This state gives the desired correlations via Eq. \eqref{eq:cut_assignment} for all $\boldsymbol{x}\in{-1,1}^n$, which finishes the proof. It implies that it is always possible to encode any bit string by taking correlators of magnitudes $1/m$. 

To end up with, we note that the state in Eq. \eqref{eq:mixed_state} is mixed. However, it can always be purified if one allows $n$ extra qubits.
In any case, we stress that the construction above is just a particular choice of valid states, giving only a lower bound to the necessary value of the correlator magnitudes in general. In fact, for the pure states obtained variationally in the main text, the correlator magnitudes we observe are significantly higher than $1/m$ (see for instance Fig. \ref{fig:histogram_distributions}).

\section*{Choice of $\alpha$}
\label{ssec:hyperparameters}

 We studied the behavior of the rescaling parameter $\alpha$ by looking at the average approximation ratios achieved at the end of the optimization for 250 random graph instances of increasing size, generated in the standard way detailed in \nameref{ssec:numdetails}. The value of $\alpha$ was increased until a plateau in solution quality was reached (see Fig. \ref{fig:tuning_alpha}). 
Based on this analysis, we fine-tune $\alpha$ for the solver at each compression rate $k$ so as to maximize $r$. 
We observe that its optimal value scales as $\alpha=\mathcal{O}(n^{\lfloor k/2 \rfloor })$,
For $k=2$, this coincides with the scaling of $1/\gamma$ analytically derived in \nameref{ssec:frustration}.

\section*{
Classical analogues of our algorithm}
\label{ssec:time_complexity}

There is a natural approach to classically mimic the algorithmic pipeline of our solver.
This consists of substituting the quantum circuit on $n$ qubits by a classical generative neural network that 
samples from a probability distribution $P$ over, for instance, $3n$ bits ($n$ bits for each of the three mutually-commuting sets in $\Pi^{(k)}$). 
With this, one can encode the binary variables into classical correlations across $k$ bits, described by $k$-body marginal distributions of $P$. 
With samples from $P$, one can Monte-Carlo-estimate all $m$ 
$k$-body correlations efficiently in the same fashion as we do with measurements on the quantum circuit.
Then, one can train the network so that 
the estimated correlations minimize a loss function analogous to that in Eq. \eqref{eq:LPi}.
However, benchmarking our scheme against the numerous classical generative neural models
is beyond the scope of the current work. 

Still, our algorithm has promising prospects
, since brickwork quantum circuits of polynomial depth in the qubit number
produce 
$k$-body expectation values that cannot be efficiently simulated classically
. An interesting exploration for future work is the connection between the amount of entanglement in the quantum circuit and the solver's performance. This could for instance be studied via classical simulations based on tensor networks \cite{patti2021tensorlyquantum} with limited bond dimensions.

\section*{Training complexity}

\begin{figure}[h!]
\centering
    \includegraphics[width=1
\linewidth]{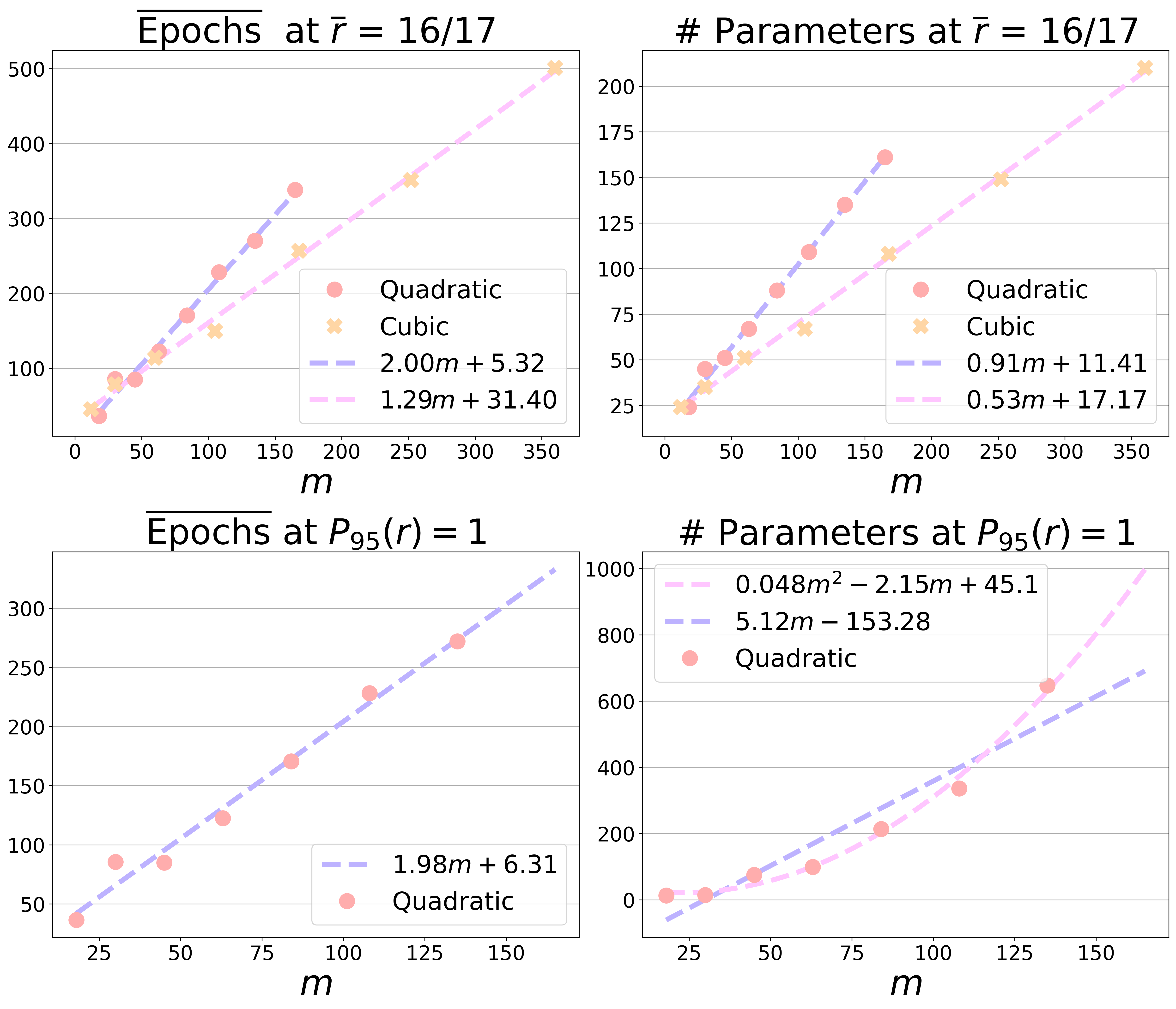}  
    \caption{{\bf Training complexity.}  Upper left: number of epochs needed to reach optimization convergence with an average approximation ratio $\overline{r}\ge 16/17\approx 0.941$ (over 250 random MaxCut instances and 5 random initializations per instance) without the local bit-swap search (quantum-circuit's output $\boldsymbol{x}$ alone) versus problem size $m$, both for quadratic and cubic compressions. A linear scaling was observed in both cases; 
    Upper right: scaling of the number of parameters in the same set up of upper left panel. 
    Lower left: observed number of epochs under the requirement that the $95^\text{th}$ percentile of $r$ was equal to one ($P_{95}(r)=1$) and including the final local search step (full solver's output $\x^*$). 
    Lower right: scaling of the number of parameters in the same set up of lower left panel. 
    }
    \label{fig:epochs_and_parameters_scaling}
\end{figure}

Here we provide further numerical details of the number of optimization parameters and training epochs for compressions of degree $k=2$ and $k=3$. 
In Fig.~\ref{fig:epochs_and_parameters_scaling} we show the observed scaling with $m$ of these figures of merit over random MaxCut instances at two different target solution qualities: $\overline{r}=0.941$ excluding the final local search step, as in Fig.~\ref{fig:numerics} (upper left and right panels); and $P_{95}(r)=1$, i.e. the $95^\text{th}$ percentile of $r$ was equal to one, now including the final local search step and increasing circuits depths accordingly (lower left and right panels). 
The latter was done to give an idea of the resources required to observe the exact solution with high probability in the practical average case. 
In terms of gate complexity, we observed a linear scaling in the first scenario (upper right), and a quadratic one in the second (lower right). 
The number of optimization epochs, on the other hand, was observed to scale linearly in both cases (upper left and upper right, respectively). 
\label{ssec:training_cpxt}

\section*{Sample complexity}
\label{ssec:sample}

Here we upper-bound the minimum number of measurements needed to estimate $\calL(\bt)$ at some arbitrary point $\bt$. 
More precisely, given $\varepsilon$, $\delta>0$, and a vector $\bt$ of variational parameters, consider the problem of estimating $\calL(\bt)$ up to additive precision $\varepsilon$ and with statistical confidence $1-\delta$. 

For each Pauli correlator $\expval{\Pi_i}$, $i\in[m]$, assume that, with confidence $1-\delta$, one has an unbiased estimator $\Pi_i^*$ with statistical error at most $\eta>0$, that is, 
\begin{equation}\label{eq:est_error_Pi}
    \Delta\langle \Pi_i \rangle := \langle \Pi_i \rangle - \Pi_i^* \quad\text{is such that}\quad\big|\Delta\langle \Pi_i \rangle\big| \le \eta\,.
\end{equation}
The corresponding error in the loss function is $\Delta\calL := \calL-\calL^*$, with $\calL^*$ given by \eq{LPi} computed using $\Pi_i^*$ instead of $\langle\Pi_i\rangle$. The multivariate Taylor theorem ensures that there is a $\xi\in[-1,1]^m$ such that
\begin{align}
\label{eq:delta_calL}
\Delta\calL = \sum_{i\in[m]} \frac{\partial \calL}{\partial{\langle\Pi_i\rangle}}\bigg|_{\xi}\,\Delta\langle\Pi_i \rangle\,. 
\end{align}

We next restrict to MaxCut, for which the expressions take simple forms in terms of the number of vertices $m$ and edges $|E|$, but the extension to weighted MaxCut is straightforward. 
For the loss function \eq{LPi}, using the basic inequalities $\tanh(x)\le1$ and $\frac{d}{dx}\tanh(x)=\sech^2(x)\le1$, one can show that $\big|\big(\frac{\partial \calL}{\partial\langle\Pi_i\rangle}\big|_\xi\big)\big| \le 2\alpha\big[d(i) + {2\beta\nu}/{m}\big]$ 
, where $d(i):=\sum_{j\in [m]} \!|W_{ij}|$ is the degree of vertex $i$. 
As a result, 
\begin{align}\label{eq:DeltaL}
|\Delta \calL|
& \le 2\,\eta\,\alpha\sum_{i\in[m]}\left[d(i) + \frac{2\beta\nu}{m}\right] \le \eta\,\alpha\,(6|E|+m), 
\end{align}
where the first step follows from the triangle inequality together with \eq{est_error_Pi}, while the second uses the identity $\sum_{i\in[m]}d(i)=2|E|$, 
$\nu = (2|E| + m-1)/4$ (see ~\nameref{ssec:reg}), and $\beta<1$. 

To ensure $|\Delta\calL|\le\varepsilon$ we require that
\begin{align}\label{eq:etabound}
\eta \le \frac{\varepsilon}{\alpha\,(6|E|+m)}\,.
\end{align}
The minimum number $S$ of samples needed to achieve such precision can be 
upper-bounded by standard arguments using the union bound and Hoeffding's inequality, which gives $S \le(4/\eta^2)\log(2m/\delta)$
Then, by virtue of Eq. \eq{etabound}, it suffices to take
\begin{align}
S &= \left\lfloor\frac{4\alpha^2}{\varepsilon^2}\,(6|E|+m)^2\log(\frac{2m}{\delta})\right\rfloor.
\end{align}
This is the general form of our upper bound. However, for the particular cases $k=2$ and $k=3$, $\alpha=\mathcal{O}(n^{\lfloor k/2\rfloor})=\mathcal{O}(m^{1/2})$ (see \nameref{ssec:hyperparameters}), hence $S=\mathcal{O}\big(m\,(6|E|+m)^2\log(\frac{2m}{\delta})/\varepsilon^2\big)$.
Moreover, in practice, it is often the case (as in all the instances in Table \ref{tab:instances}) that $|E|$ is linear in $m$, making the statistical overhead $\tilde{\mathcal{O}}({m^3})$.

\begin{table}[hb!]
\begin{ruledtabular}
\begin{tabular}{c@{} c|c|ccccc|c@{\hspace{-1ex}}c} 
\multirow{2}{*}{{\it Graph} } & \multirow{2}{*}{$m$} & \multirow{2}{*}{{\it Alg.} } & \multirow{2}{*}{$k$ } & \multirow{2}{*}{$n$ }   &\multirow{2}{*}{{\it 2-q} }  &\multirow{2}{*}{{\it Par.}  } &\multirow{2}{*}{{\it Runs}}   & \qquad\qquad $r$ &  
\\\cline{9-10} 
 & & &   &  &  &  & &  {\it Mean} & {\it Max} \\  \hline
\multirow{3}{*}{ {\it torus}} & \multirow{3}{*}{512}   &{\it BM}  & -   & - & -  & 512 &100  & 0.939 & 0.969  \\
 & &{\it MBE}  & 1   &256 & 768  & 1792 &30  & 0.948 & 0.978  \\ 
 &  & {\it Our} & 4   &\textbf{10} & \textbf{125}   & \textbf{510}  & 30 & \textbf{0.961} & \textbf{0.987} \\
  \cline{1-10}
\multirow{2}{*}{{\it G14}}  &\multirow{2}{*}{800} &{\it BM}&-&-&-&800&1&0.984&-\\
& &{\it Our}&5&11&200&811&10&0.985&0.991\\ \cline{1-10}
\multirow{2}{*}{{\it G23}} &\multirow{2}{*}{2000}  &{\it BM}&-&-&-&2000&1&0.989&-\\
& &{\it Our}&6&12&498&2004&10&0.992&0.995\\ \cline{1-10}
\multirow{2}{*}{{\it G60}} &\multirow{2}{*}{7000}&{\it BM}&-&-&-&7000&1&0.970&-\\
& &{\it Our}&5&15&1750&7014&5&0.975&0.978\\
\end{tabular}
\end{ruledtabular}
\caption{ {\bf Single-shot resources and quality of solutions.} For each instance we display, for each heuristic, $m$, $k$, $n$, the 2-qubit gate counts, the number of optimization parameters used during training, and the number of random initializations. The last two columns report the average approximation ratios and the best one observed at the end of training.
 In the case of {\it pm3-8-80} ({\it torus}), for which bigger statistics are available, we observed significant improvements in the amount of required resources compared to the single-qubit multi-basis encoding of \cite{Patti2022multibasis}, which is equivalent to our PCE with $k=1$. Additionally, both the average and peak approximation ratios showed improvement over both {\it MBE} and {\it BM}. For the remaining three instances, where data from only a single initialization is available for the BM algorithm, we noted the average performance of our method to be higher.} \label{tab:comparison}
\end{table} 
\section*{Details on the comparison with Burer-Monteiro}
\label{ssec:extra_numerics}
Here we provide more detailed information on the results of our method on the benchmark instances reported in Tab. \ref{tab:instances}. 
In their paper, Burer and Monteiro provide an effective method to carry out their non-convex optimization (see Algorithm-1 in Ref.\cite{Burer_Monteiro_2003}). After reaching a local minimum, and extensive local search is performed (one- and two-bit swap search). Then, the obtained parameters are perturbed, and a new minimization is carried out until convergence is reached. If a local search on the new solution leads to a better value of the cut, the parameters are updated, and the procedure repeated. If, after $N$ perturbations, no better cut is found, the optimization is halted. 
For all the benchmarked instances, they provide the results obtained with different choices of number of initializations and of $N$.  
On the other hand, in our method, we execute a single (quantum) optimization followed by a final local search, which effectively places it on equal footing to the BM algorithm with $N=0$.  Given that, in the table (\ref{tab:comparison}) we provide a comparison of the approximation ratios with the single-shot version of BM.

\section*{Comparison between experimental and naive solutions}
\label{ssec:random}
\begin{figure}[]
\centering
    \includegraphics[width=0.9\linewidth]{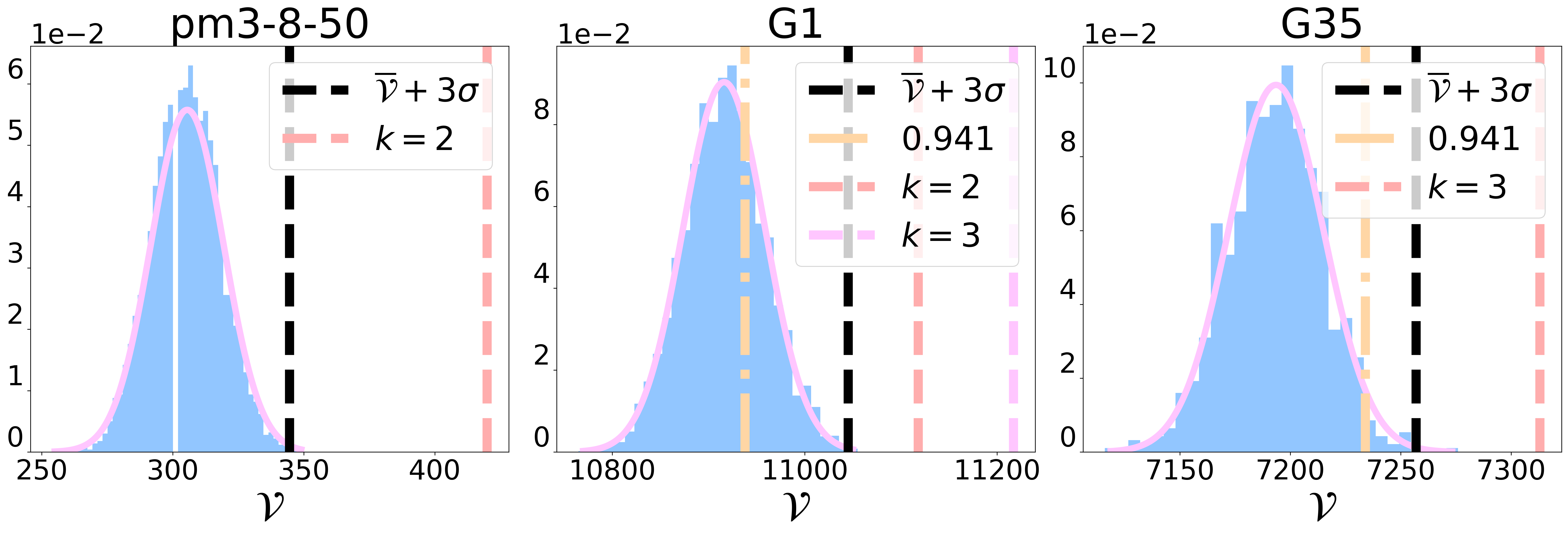}  
    \caption{{\bf Experimental versus naive solutions.}  Cut value distributions of \lq\lq naive\rq\rq~solutions obtained by performing local search over a randomly picked bitstring ($1000$ random cut assignments were used) for the three benchmarking instances used in the experiment (see Table \ref{tab:instances}). Each plot shows the $3\sigma$ right tail of the distribution based on a Gaussian fit (pink curve), the hardness bound $0.941$, as well as the observed experimental results with $k=2$ and $k=3$. Clearly, the experimental solution is non-trivial with respect the naive ones. 
    }
    \label{fig:histogram_experiment}
\end{figure}

Here we compare the solutions found in our experimental demonstrations to \lq\lq naive\rq\rq~solutions obtained by randomly picking a graph partition and performing a local search over it. 
Fig. \ref{fig:histogram_experiment} shows the cut value distributions over 1000 naive solutions together with the cut value of our experimental solutions. The distribution appears Gaussian, as indicated by the Gaussian fit (pink curve). The vertical lines locate the $3\sigma$ right tail, the hardness threshold $0.941$, and our experimental solutions for the $k=2$ and $k=3$ encodings. The results clearly indicate the non-trivial character of our solutions, which lie beyond $3\sigma$ for the first two instances and near $3\sigma$ for the last one.  
We recall that the hardness bound is not shown for {\it pm3-8-50} (left plot) since this is a weighted MaxCut instance. 

\section*{Approximate parent Hamiltonian}
\label{ssec:parent}
Here, we show that it is possible to construct a parent Hamiltonian for approximate MaxCut solutions via Pauli-correlation encoding retaining the polynomial compression of our method. Our construction closely follows the footprints of \textbf{Proposition 1} of Ref. \cite{fuller2021approximate}. 
With it, we can show the following:

\textit{Given a weighted graph $G = (V, E)$ of degree $\deg(G)$ and $|V| = m$, there exists a map $\varrho$ from bit strings $\boldsymbol{x}\in\{-1, 1\}^m$ 
to density matrices $\varrho(\boldsymbol{x})$, and a Hamiltonian,
 \begin{equation}\label{eq:parent_h}
     H = \sum_{e \in E} \frac{1}{2}(I- 
     \frac{1}{\gamma^2} O_e),
\end{equation}
on $n = \mathcal{O}(\text{deg}(G)\, m^{\frac{1}{k}})$ qubits, with $k$ an integer of our choice, $O_e$ a $2k$-body Pauli string, and 
$\gamma$ a suitable constant [see Eq. \eqref{eq:pi_trace_parent_h}] 
such that
 \begin{equation}
     \text{Tr}[H \cdot \varrho(\boldsymbol{x})]=\mathcal{V}(\boldsymbol{x}),
 \end{equation}
for all $\boldsymbol{x} \in \{-1, 1\}^m$. Moreover, the construction of $H$ has time complexity $\mathcal{O}(m \log(m) + m \deg(G))$.}

The first step to building $H$ requires us to color the graph.  We call a partition $\{V_c\}_{c \in [C]}$ a coloring of the graph $G = (V, E)$ into $C$ colors, if for every edge  $e_{i, j} \in E$, connecting vertices $i$ and $j$, we have that $i \in V_c$ and $j \in V_{c'}$ for $c \neq c'$, i.e, vertices with the same color are guaranteed not to be connected by any edge. From now on we label the vertices such that $(\lambda, c) \in V_c$ is the $\lambda$-th element of color $c$. The basic idea is then to assign a different group of qubits to each color and apply our compression scheme to each color independently. Note that we could even choose a different compression rate $k_c$ for each color, since each sub-partition will in general have a different number of vertices. However, we choose $k_c = k$ for all colors for simplicity. 

As discussed in the main text, we can encode the $m_c = |V_{c}|$ vertices in each color $c$ using $n_c = \mathcal{O}(|V_c|^{1/k})$ qubits. 
That is, we choose $C$ sets of $k$-body Pauli strings, $\Pi_{c} = \{\Pi_{\lambda, c}\}_{\lambda\in [m_c]}$, with support in $n_c$ qubits, 
and use a \textit{Pauli-correlation encoding} with respect to $\Pi = \cup_{c} \Pi_c$.
This, since $|V_c| \leq m$, gives a total number of qubits
\begin{equation}
    n = \sum_{c \in [C]}{n_c} = \mathcal{O}(C\, m^{1/k}).
\end{equation}
Using the large-degree-first algorithm from \cite{Welsh_67}, one can find a coloring of a graph with $C=\mathcal{O}(\deg(G))$ in time $\mathcal{O}(m \log(m) + m \deg(G))$, which gives the promised scaling $n = \mathcal{O}(\text{deg}(G)m^{\frac{1}{k}})$.

Now, let us define $\varrho (\boldsymbol{x})$ to be a state such that
\begin{equation}\label{eq:pi_trace_parent_h}
    \text{Tr}[\Pi_{\lambda, c} \,\varrho(\boldsymbol{x})] = 
    \gamma\, x_{\lambda, c},
\end{equation}
where $x_{\lambda, c}$ is the 
component of $\boldsymbol{x}$ associated with vertex $(\lambda, c)$ 
and 
$\gamma$ a small-enough constant to guarantee that  $\varrho(\boldsymbol{x})$ is a valid state, as discussed in \nameref{ssec:frustration}).
In addition, take each $O_e$ appearing in Eq. \eqref{eq:parent_h} as 
$O_e = \Pi_{\lambda,c}\,\Pi_{\nu,c'}$, with $(\lambda,c)$ and $(\nu,c')$ the two nodes connected by edge $e$. 
Due to the coloring of the graph, we know that $c \neq c'$. This, in turn, due to the assignment of different qubits to each color, guarantees that  $\Pi_{\lambda,c}$ and $\Pi_{\nu,c'}$ have non overlapping support.
This, together with Eq. \eqref{eq:pi_trace_parent_h}, implies
\begin{equation}\label{eq:o_trace_parent_h}
    \text{Tr}[O_e\, \varrho(\boldsymbol{x})] = \text{Tr}[\Pi_{\lambda,c}\,\Pi_{\nu,c'} \varrho(\boldsymbol{x})] = 
    \gamma^2 x_{\lambda,c}\,\,x_{\nu,c'}.
\end{equation}

Finally, Eqs. \eqref{eq:parent_h} and \eqref{eq:o_trace_parent_h} together give
\begin{equation}\label{eq:parent_h_cut}
    \text{Tr}[H \cdot \varrho(\boldsymbol{x})]=\mathcal{V}(\boldsymbol{x}).
\end{equation}

Equation \eqref{eq:parent_h_cut} shows us that the state $\varrho(\boldsymbol{x}_{max})$ with maximum energy over the image of the map $\varrho$ is associated with the solution to our problem, specifically $\mathcal{V}_{max} = \mathcal{V}(\boldsymbol{x}_{max})$. This tells us that by solving for the ground state of $-H$, we can get an approximate solution to the MaxCut problem in question. This solution will only be approximate because the ground state $\varrho_\text{min}$ of $-H$ is not in general in the image of $\varrho$, i.e. $\text{Tr}(-H \varrho_\text{min}) \leq \min_{\boldsymbol{x}}\text{Tr}(-H\varrho(\boldsymbol{x}))$. We also note that

\begin{equation}
    \underset{\boldsymbol{x}}{\text{argmax}} \, \text{Tr}(H\varrho(\boldsymbol{x})) = \underset{\boldsymbol{x}}{\text{argmin}}  \, \text{Tr}\left[ \left(\sum_{e \in E}O_e\right)  \varrho(\boldsymbol{x})\right].
\end{equation}
Interestingly, this implies that 
$\gamma$ (or any other hyper-parameter in Eq. \eqref{eq:LPi}) is not necessary to find the solution bit-string; we may take any value of 
$\gamma$ in Eq. \eqref{eq:parent_h}. The specific choice of 
$\gamma$ is needed only to match the corresponding cut values, not for the string itself.

All in all, however, this approach comes with two caveats. First, as evident from the last expression, the qubit-number compression is restricted by the connectivity of the graph. For instance, in the limiting case of fully-connected graphs, no compression is possible (even though heuristic graph-sparsification techniques may mitigate this problem). Secondly, in Ref. \cite{teramoto2023quantumrelaxation}, analytical lower bounds to the approximation ratios were derived that decrease with the compression rate. This is consistent with the intuition that too high compression rates can compromise the quality of the solution.
Nevertheless, for graphs with restricted connectivity, having access to a parent Hamiltonian opens up interesting opportunities. For instance, QAOA-type approaches \cite{Farhi2014} may be combined with our variational solver, the former preparing approximate solution states (pre-training) and the latter refining them.

\section*{Analytical barren plateau characterization}
\label{ssec:bp_analytics}
In this section, we analytically compute the variance of our loss function for deep circuits. That is, we compute the value to which the variance converges as the circuit depth increases.
The nonlinear hyperbolic tangent appearing in the loss renders an exact computation involved. Hence, we begin by computing the variance for the simplified quadratic loss function $\LC^{(\rm qua)}$ analyzed in Figure \ref{fig:histogram_distributions}.
This simplified calculation will be instrumental to obtain the variance of the actual loss function. More precisely, we first show, under the assumption that the circuit ensemble under random parameter initializations is a 4-design over the special unitary group, that the variance of $\LC^{(\rm qua)}$ is equal to $\frac{1}{d^2}\sum_{(i,j)\in E} w_{ij}^2   + \OC\left(\frac{1}{d^4}\right)$, where $d=2^n$ is the dimension of the Hilbert space. Then, we show, under the assumption that the circuit is fully Haar random, that the variance of $\LC$ is given by $\frac{\alpha^4}{d^2}\sum_{(i.j)\in E} w_{ij}^2   + \OC\left(\frac{\alpha^6}{d^3}\right)$.

The simplified loss function has the form
\small
\begin{equation}
    \LC^{(\rm qua)} = \sum_{(i,j)\in E} w_{ij}\Tr\left[U(\thv)\,\varrho\, U(\thv)^\dagger \Pi_i\right] \Tr\left[U(\thv)\,\varrho \,U(\thv)^\dagger \Pi_j\right]\,,
\end{equation}
\normalsize
with $\varrho$ a pure state.
For arbitrary depths, one would be interested in computing the variance ${\rm Var}_\thv\left[\LC^{(\rm qua)}\right]$ of $\LC^{(\rm qua)}$ over a uniform sampling of parameter values in the interval $[0,2\pi]$. However, such computation is non-trivial. Instead, we will resort to representation-theoretic techniques and compute the variance ${\rm Var}_{\mathbb{SU}(d)}\left(\LC^{(\rm qua)}\right)$ assuming that the quantum circuit is a design over the special unitary group, which we denote as $\mathbb{SU}(d)$. In practice, if the circuit is deep enough it will always form a design over the dynamical Lie group associated to the circuit's generators~\cite{ragone2023unified}, thus justifying the utility of the computation. When the circuit's generators are traceless and universal (which is our case), the corresponding dynamical Lie group is $\mathbb{SU}(d)$. 
Since we will work at the special unitary group level, we will henceforth drop the explicit dependence of the unitaries on the variational parameters $\thv$. 

We start by computing
\onecolumngrid
\begin{equation} 
    \mathbb{E}_{\mathbb{SU}(d)}\left[\left(\LC^{(\rm qua)}\right)^2\right]= \sum_{(i,j)\in E} \sum_{(k,l)\in E} w_{ij}w_{kl}\int d\mu(U)\Tr\left[U^{\otimes 4}\varrho^{\otimes 4} (U^\dagger)^{\otimes 4}\, \Pi_i \otimes \Pi_j\otimes \Pi_k\otimes \Pi_l\right]\,,
\end{equation} 
where $d\mu(U)$ is the volume element from the Haar measure, and we used the property that $\Tr[A\otimes B] = \Tr[A]\Tr[B]$. (In fact, for the simplified loss functions it suffices that $d\mu(U)$ defines a 4-design, the fully-random Haar measure will be needed only for the actual loss function below.)
Using standard Weingarten calculus techniques~\cite{garcia2023deep}, we have that
\begin{equation}\begin{split}
    &\mathbb{E}_{\mathbb{SU}(d)}\left[\Tr\left[U^{\otimes 4} \varrho^{\otimes 4}  (U^\dagger)^{ \otimes 4}\, \Pi_i \otimes \Pi_j\otimes \Pi_k\otimes \Pi_l\right] \right] =\\ &\frac{1}{d^4}\sum_{\sigma\in S_4}\Tr[\varrho^{\otimes 4} P_d(\sigma)]\Tr[P_d(\sigma^{-1})\,\Pi_i \otimes \Pi_j\otimes \Pi_k\otimes \Pi_l]  +\frac{1}{d^4}\sum_{\sigma,\pi\in S_4}c_{\sigma,\pi}\Tr[\varrho^{\otimes 4} P_d(\sigma)]\Tr[P_d(\pi) \Pi_i \otimes \Pi_j\otimes \Pi_k\otimes \Pi_l] =\\  &\frac{1}{d^4}\sum_{\sigma\in S_4} \Tr[P_d(\sigma^{-1})\,\Pi_i \otimes \Pi_j\otimes \Pi_k\otimes \Pi_l] +\frac{1}{d^4}\sum_{\sigma,\pi\in S_4}c_{\sigma,\pi}\Tr[P_d(\pi) \Pi_i \otimes \Pi_j\otimes \Pi_k\otimes \Pi_l] \,,
\end{split}
\end{equation}
where $d=2^n$ is the dimension of the Hilbert space, $c_{\sigma,\pi}\in\OC(1/d)$, and $P_d$ is the  representation of the Symmetric group $\mathbb{S}_t$ that permutes the $d$-dimensional subsystems in the $t$-fold tensor product
Hilbert space, $\HC^{\otimes t}$ (i.e. $P_d(\sigma)= \sum_{i_1,\dots,i_t=0}^{d-1} |i_{\sigma^{-1}(1)},\dots,i_{\sigma^{-1}(t)} \rangle\langle i_1,\dots,i_t|$, for a permutation $\sigma\in \mathbb{S}_t$). Furthermore, we used that $\Tr[\varrho^{\otimes 4} P_d(\sigma)]=1$ $\forall \sigma\in \mathbb{S}_4$ since $\varrho$ is pure. 

Let us now take a look at the permutations in $\mathbb{S}_4$. 
Using cycle notation (see, e.g., Supp. Info. C of \cite{garcia2023deep}), the $4!=24$ permutations can be classified as follows: 
the identity, six transpositions, three double-transpositions, eight 3-cycles and six 4-cycles. We now note that for any $\sigma$ containing an odd-length cycle, the term $\Tr[P_d(\sigma) \Pi_i \otimes \Pi_j\otimes \Pi_k\otimes \Pi_l] $ vanishes, since $\Pi_i, \Pi_j, \Pi_k, \Pi_l$ are all traceless. Hence, we are left with the double transpositions and the 4-cycles. 
For the double transpositions $(ik)(jl)$ and $(il)(jk)$ we find that $\Tr[P_d(\sigma) \Pi_i \otimes \Pi_j\otimes \Pi_k\otimes \Pi_l]$ is equal to $ d^2 \delta_{\Pi_i\Pi_k} \delta_{\Pi_j\Pi_l}$ and $d^2 \delta_{\Pi_i\Pi_l} \delta_{\Pi_j\Pi_k}$, respectively, while for $(ij)(kl)$ we have $\Tr[P_d(\sigma) \Pi_i \otimes \Pi_j\otimes \Pi_k\otimes \Pi_l] = 0$ since $\Pi_i\neq \Pi_j$. 
Noting that $\delta_{\Pi_i\Pi_k}=\delta_{ik}$, we thus arrive at
   \begin{equation}
    \mathbb{E}_{\mathbb{SU}(d)}\left[\left(\LC^{(\rm qua)}\right)^2\right]= \frac{1}{d^2}\sum_{(i,j)\in E} \sum_{(k,l)\in E} w_{ij}w_{kl} \,\delta_{ik} \delta_{jl} + \OC\left(\frac{1}{d^4}\right)= \frac{1}{d^2}\sum_{(i.j)\in E} w_{ij}^2  + \OC\left(\frac{1}{d^4}\right) \,,
\end{equation} 
where the terms in $\OC\left(\frac{1}{d^4}\right)$ account for the 4-cycles contributions.
On the other hand, we have
\small
\begin{equation} 
    \mathbb{E}_{\mathbb{SU}(d)}\left[\LC^{(\rm qua)}\right]= \sum_{(i,j)\in E} w_{ij}\left(\frac{1}{d^2}\sum_{\sigma\in S_2}\Tr[\varrho^{\otimes 2} P_d(\sigma)]\Tr[P_d(\sigma^{-1})\,\Pi_i \otimes \Pi_j]  +\frac{1}{d^2}\sum_{\sigma,\pi\in S_2}c_{\sigma,\pi}\Tr[\varrho^{\otimes 2} P_d(\sigma)]\Tr[P_d(\pi) \Pi_i \otimes \Pi_j ]\right) =0 \,,
\end{equation}
\normalsize
where we used that $\Tr[\Pi_i \otimes \Pi_j ]=\Tr\left[{\rm SWAP}\, \Pi_i \otimes \Pi_j \right]=0$. The variance therefore reads 
\begin{equation} \label{eq-appv:variance}
    {\rm Var}_{\mathbb{SU}(d)}\left(\LC^{(\rm qua)}\right) = \frac{1}{d^2}\sum_{(i,j)\in E} w_{ij}^2   + \OC\left(\frac{1}{d^4}\right)\,.
\end{equation}
In particular, for the unweighted version of the MaxCut problem, the variance is given by ${\rm Var}_{\mathbb{SU}(d)}\left(\LC^{(\rm qua)}\right) = \frac{|E|}{d^2}$. Equation~\eqref{eq-appv:variance} implies that if the quantum circuit is a $4$-design over the special unitary group, then the variance of the loss function is exponentially suppressed as $\OC(1/2^{2n})$.

We now compute the variance for the actual loss function in the main text, Eq. \eq{LPi}, namely
\begin{align}
    \LC  &= \sum_{(i,j)\in E} w_{ij} \tanh\left(\alpha\Tr\left[U(\thv)\,\varrho\, U(\thv)^\dagger \Pi_i\right]\right) \tanh\left(\alpha\Tr\left[U(\thv)\,\varrho\, U(\thv)^\dagger \Pi_j\right]\right) + \beta\,\nu \left[\frac{1}{m}\sum_{i\in V} \tanh\left(\alpha\Tr\left[U(\thv)\,\varrho\, U(\thv)^\dagger \Pi_i\right]\right)^2\right]^2 \notag\\
    &\equiv \LC^{({\tanh})} +\LC^{({\rm reg})}\,. 
\end{align}
We proceed by using the Taylor-series expansion of the hyperbolic tangent, namely $\tanh(x) = \sum_{s=1}^\infty C_s\, \,x^{2s-1}$, where $C_s\equiv \frac{2^{2s}(2^{2s}-1) B_{2s}}{(2s)!}$ and $B_{2s}$ are the Bernoulli numbers.
We start by computing 
\small
    \begin{equation} 
    \label{eq:coefs_C} 
    \begin{split}
    \mathbb{E}_{\mathbb{SU}(d)}\left[\left(\LC^{({\tanh})}\right)^2\right] = \sum_{(i,j)\in E} \sum_{(k,l)\in E} w_{ij}w_{kl} \sum_{s_1,s_2,s_3,s_4=1}^\infty C_{s_1} C_{s_2} C_{s_3} C_{s_4}\,\int d\mu(U)&\left(\alpha\Tr\left[U\,\varrho\, U^\dagger \Pi_i\right]\right)^{2s_1-1} \left(\alpha\Tr\left[U\,\varrho\, U^\dagger \Pi_j\right]\right)^{2s_2-1}\\ &\left(\alpha\Tr\left[U\,\varrho\, U^\dagger \Pi_k\right]\right)^{2s_3-1} \left(\alpha\Tr\left[U\,\varrho \,U^\dagger \Pi_l\right]\right)^{2s_4-1}\,. \end{split}
\end{equation}
\normalsize
Here, we will be dealing with quantities of the form
\begin{equation}
     \alpha^t\int d\mu(U)\Tr\left[U(\thv)^{\otimes t} \varrho^{\otimes t} \left(U(\thv)^\dagger\right)^{\otimes t}\, \Pi_i^{\otimes t_1} \otimes \Pi_j^{\otimes t_2} \otimes\Pi_k^{\otimes t_3}\otimes \Pi_l^{\otimes t_4}\right]\,,
\end{equation}
where $t_\gamma =2s_\gamma-1$ and $t=t_1+t_2+t_3+t_4$. Using asymptotic Weingarten calculus (see, e.g., Ref.~\cite{garcia2023deep}), it can be shown that 
\small
\begin{equation}\label{eq-app:integral}\begin{split}
      \int d\mu(U)\Tr\left[U^{\otimes t} \varrho^{\otimes t} \left(U^\dagger\right)^{\otimes t}\, \Pi_i^{\otimes t_1} \otimes \Pi_j^{\otimes t_2} \otimes\Pi_k^{\otimes t_3}\otimes \Pi_l^{\otimes t_4}\right] &= \frac{\left|\TC_{t_1+t_3}\right| \delta_{ik} + \left|\TC_{t_2+t_4}\right|\delta_{jl}}{d^{t/2}}\\&+\frac{(1-\delta_{ik})(\left|\TC_{t_1}\right|+\left|\TC_{t_3}\right|)+(1-\delta_{jl})(\left|\TC_{t_2}\right|+\left|\TC_{t_4}\right|)}{d^{t/2}}+\OC\left(\frac{1}{d^{t/2+1}}\right)\,,\end{split}
\end{equation}
\normalsize
where $\left|\TC_{\ell}\right|\equiv \frac{\ell!}{2^{\ell/2}\left(\ell/2\right)!}$ denotes the number of permutations that consist of exactly $\ell$ disjoint transpositions.  
The dimensional dependence obtained in Eq. \eqref{eq-app:integral} implies that, for large-enough $d$, Eq. \eqref{eq:coefs_C} is well-approximated by its first-order terms, i.e. the terms where $t=4$. These first-order terms lead to a total contribution $\OC\left(1/d^2\right)$, while the rest contribute with $\OC\left(1/d^3\right)$. More precisely, we find that
\begin{equation} \label{eq-app:var-tan}
    \mathbb{E}_{\mathbb{SU}(d)}\left[\left(\LC^{({\tanh})}\right)^2\right] =  \frac{\alpha^4}{d^2}\sum_{(i.j)\in E} w_{ij}^2   + \OC\left(\frac{\alpha^6}{d^3}\right) \,.
\end{equation}
Furthermore, since $\Tr[P_d(\sigma)\,\Pi_i^{\otimes t_1} \otimes \Pi_j^{\otimes t_2} ]=0$ $\forall \sigma \in\mathbb{S}_{t_1+t_2}$ (this is true because the hyperbolic tangent is an odd function, which implies that $t_1$ and $t_2$ are odd),
it follows that  $\mathbb{E}_{\mathbb{SU}(d)}\left[\LC^{({\tanh})}\right]=0\,.$

Finally, it remains to include the contributions to the variance coming from the regularization term $\LC^{(\rm reg)}$. 
Here, it suffices to notice that 
$\mathbb{E}_{\mathbb{SU}(d)}\left[\left(\LC^{({\rm reg})}\right)^2 \right] \in \OC\left(\frac{1}{d^4}\right)\,,
$
$\mathbb{E}_{\mathbb{SU}(d)}\left[\LC^{(\rm reg)}\LC^{({\tanh})} \right] \in \OC\left(\frac{1}{d^3}\right)\,,$ and $\mathbb{E}_{\mathbb{SU}(d)}\left[\LC^{(\rm reg)} \right]^2 \in \OC\left(\frac{1}{d^4}\right)\,,$ which follow from applying Eq.~\eqref{eq-app:integral}. 
Putting all together, the final result is
\begin{equation}
\label{eq:plateau_tanh}
      {\rm Var}_{\mathbb{SU}(d)}\left(\LC\right) =  \frac{\alpha^4}{d^2}\sum_{(i.j)\in E} w_{ij}^2   + \OC\left(\frac{\alpha^6}{d^3}\right) \,.
\end{equation}
We remark here that since the hyperbolic tangent is expanded as an infinite Taylor series, we require the circuit to be fully Haar random under the parameters initialization in order for Eq.~\eqref{eq:plateau_tanh} to hold exactly. However, if the circuit ensemble is already a $4$-design, we do not expect higher order terms to differ significantly but rather to become negligible as $n\rightarrow\infty$.

\end{document}